\documentclass[10pt,journal,compsoc]{IEEEtran}
\ifCLASSOPTIONcompsoc
  \usepackage[nocompress]{cite}
\else
  \usepackage{cite}
\fi
\ifCLASSINFOpdf
\else
\fi
\usepackage{amsmath}
\usepackage{algorithmic}
\usepackage{url}

\usepackage{xspace}
\usepackage{graphicx}
\usepackage{subfiles} 
\usepackage{multirow}

\usepackage[ruled,linesnumbered]{algorithm2e}
\usepackage{listings}
\usepackage{subcaption}
\usepackage{subfloat}
\usepackage{color,xcolor,colortbl}
\usepackage{xspace}
\usepackage{enumitem}
\usepackage{xr}
\usepackage{url}
\usepackage{booktabs}
\usepackage{diagbox}
\usepackage{array}
\usepackage{cite}
\usepackage{amssymb}
\usepackage{tcolorbox}

\usepackage[]{collab}
\collabAuthor{yc}{purple}{Yichen'Note}
\definecolor{codegreen}{rgb}{0,0.6,0}
\definecolor{codegray}{rgb}{0.5,0.5,0.5}
\definecolor{codepurple}{rgb}{0.58,0,0.82}
\definecolor{backcolour}{rgb}{0.95,0.95,0.92}
\definecolor{viol}{RGB}{134,0,175}

\collabAuthor{yjx}{blue}{Ye}

\newcommand{\tool}{MAGNET\xspace}
\newcommand{\gnn}{MHAGNN\xspace}

\newcommand{\vd}{Top-CWE\xspace}

\newcommand{\mpa}{meta-path attention\xspace}

\newcommand{\eg}{\textit{e}.\textit{g}.,\xspace}
\newcommand{\ie}{i.e.,\xspace}
\newcommand{\et}{\textit{et} \textit{al.}\xspace}
\newcommand{\etc}{\textit{etc.}\xspace}

\newcommand{\ivd}{IVDetect\xspace}
\newcommand{\tabincell}[2]{\begin{tabular}{@{}#1@{}}#2\end{tabular}}
\newtheorem{myDef}{Definition}
\newcommand{\http}{\url{https://github.com/xmwenxincheng/MAGNET}}

\begin{document}
%


\title{Meta-Path Based Attentional Graph Learning Model for Vulnerability Detection}

%
%
%
%

\author{\IEEEauthorblockN{Xin-Cheng Wen\IEEEauthorrefmark{2},
Cuiyun Gao\IEEEauthorrefmark{1}\thanks{* corresponding author.}\IEEEauthorrefmark{2}\IEEEauthorrefmark{3},
Jiaxin Ye\IEEEauthorrefmark{4},
Yichen Li\IEEEauthorrefmark{5},
Zhihong Tian\IEEEauthorrefmark{6},
Yan Jia\IEEEauthorrefmark{7},
and Xuan Wang\IEEEauthorrefmark{2}\IEEEauthorrefmark{3}
} \\
\IEEEauthorblockA{\IEEEauthorrefmark{2}}Harbin Institute of Technology, Shenzhen, China\\
\IEEEauthorrefmark{3} Guangdong Key Laboratory of New Security and Intelligence Technology, Shenzhen, China\\
\IEEEauthorrefmark{4} Fudan University, Shanghai, China\\
\IEEEauthorrefmark{5} The Chinese University of Hong Kong, Hong Kong, China\\ 
\IEEEauthorrefmark{6} GuangZhou University, GuangZhou, China\\
\IEEEauthorrefmark{7} Peng Cheng Laboratory, Shenzhen, China\\
\IEEEauthorblockA{xiamenwxc@foxmail.com, \{gaocuiyun, jiayan2020\}@hit.edu.cn, jxye22@m.fudan.edu.cn, 
ycli21@cse.cuhk.edu.hk, tianzhihong@gzhu.edu.cn, wangxuan@cs.hitsz.edu.cn}}
\IEEEtitleabstractindextext{%
\begin{abstract}

In recent years, deep learning (DL)-based methods have been widely used in code vulnerability detection. The DL-based methods typically extract structural information from source code, e.g., code structure graph, and adopt neural networks such as Graph Neural Networks (GNNs) to learn the graph representations. However, these methods fail to consider the heterogeneous relations in the code structure graph, i.e., the heterogeneous relations mean that the different types of edges connect different types of nodes in the graph, which may obstruct the graph representation learning. Besides, these methods are limited in capturing long-range dependencies due to the deep levels in the code structure graph. In this paper,  we propose a \textbf{M}eta-path based \textbf{A}ttentional \textbf{G}raph learning model for code vul\textbf{NE}rability de\textbf{T}ection, called \textbf{\tool}. \tool constructs a multi-granularity meta-path graph for each code snippet, in which the heterogeneous relations are denoted as meta-paths to represent the structural information. A meta-path based hierarchical attentional graph neural network is also proposed to capture the relations between distant nodes in the graph. We evaluate \tool on three public datasets and the results show that \tool outperforms the best baseline method in terms of F1 score by 6.32\%, 21.50\%, and 25.40\%, respectively. \tool also achieves the best performance among all the baseline methods in detecting Top-25 most dangerous Common Weakness Enumerations (CWEs), further demonstrating its effectiveness in vulnerability detection.

\end{abstract}

}

\maketitle

\IEEEdisplaynontitleabstractindextext

%
\IEEEpeerreviewmaketitle

\IEEEraisesectionheading{\section{Introduction}\label{sec:introduction}}


Software vulnerabilities are generally specific flaws or oversights in the pieces of software that allow attackers to disrupt or damage a computer system or program~\cite{def1}, leading to security risks~\cite{ill1,ill2,9778273/Tian} such as system crash and data leakage.
The ever-growing number of software vulnerabilities poses a threat to social public security. For instance, Bugcrowd~\cite{bugcrowd}, a crowdsourced security platform,
reported a 185\% increase in the number of high-risk vulnerabilities in 2021 compared to the previous year.
In December 2021, only 11 days after the Apache Log4j2's remote code execution vulnerability was disclosed, attackers exploited the vulnerability to attack Belgian network systems, causing
system outages~\cite{wikipedia}. Thus, software vulnerability detection is critical for improving the security of society.



\begin{figure*}[t]
\centering
\subfloat[A source code snippet]{\includegraphics[width=0.4\linewidth]{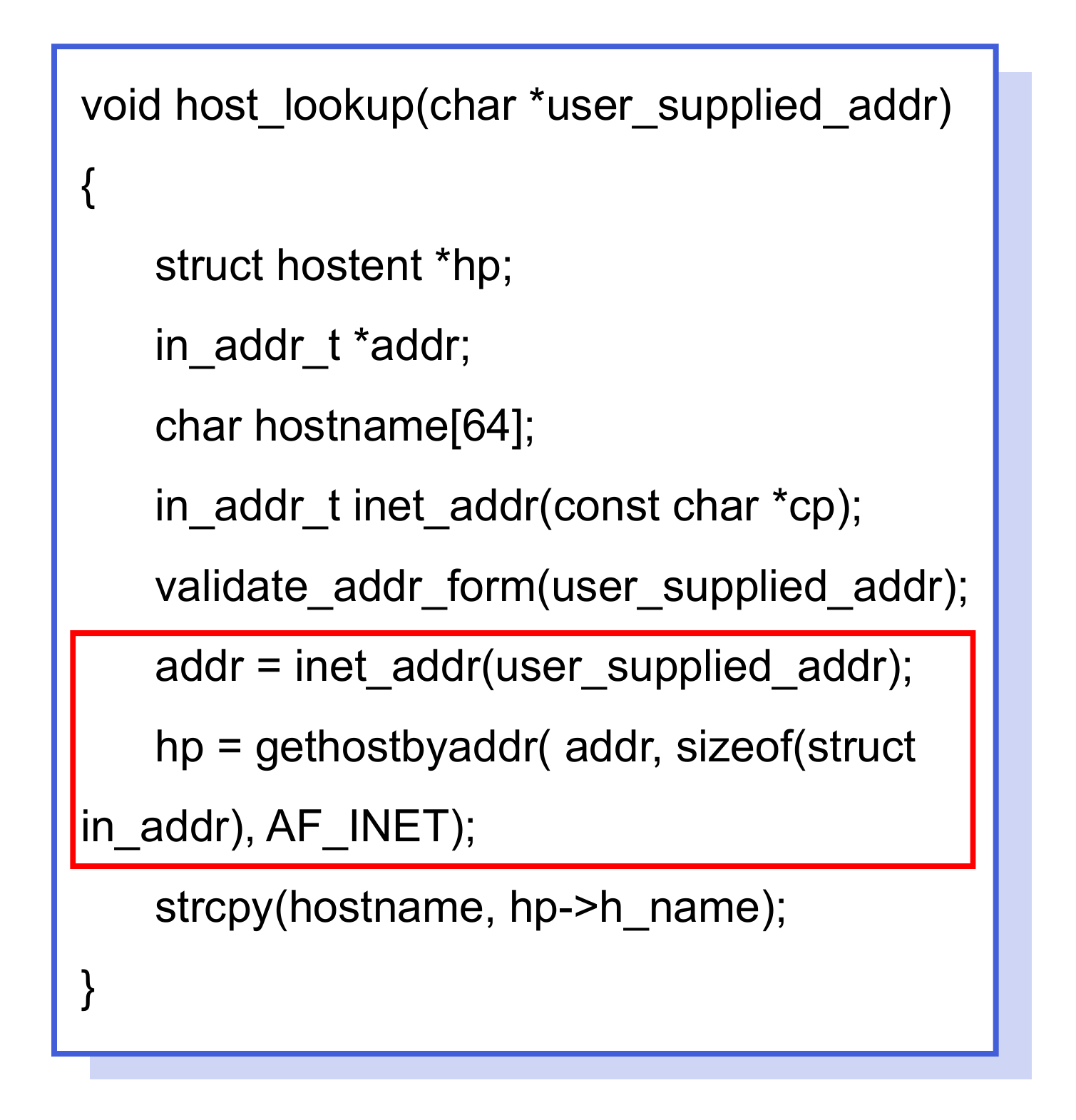}%
\label{source}}
\hfil
\subfloat[A partial code structure graph]{\includegraphics[width=0.6\linewidth]{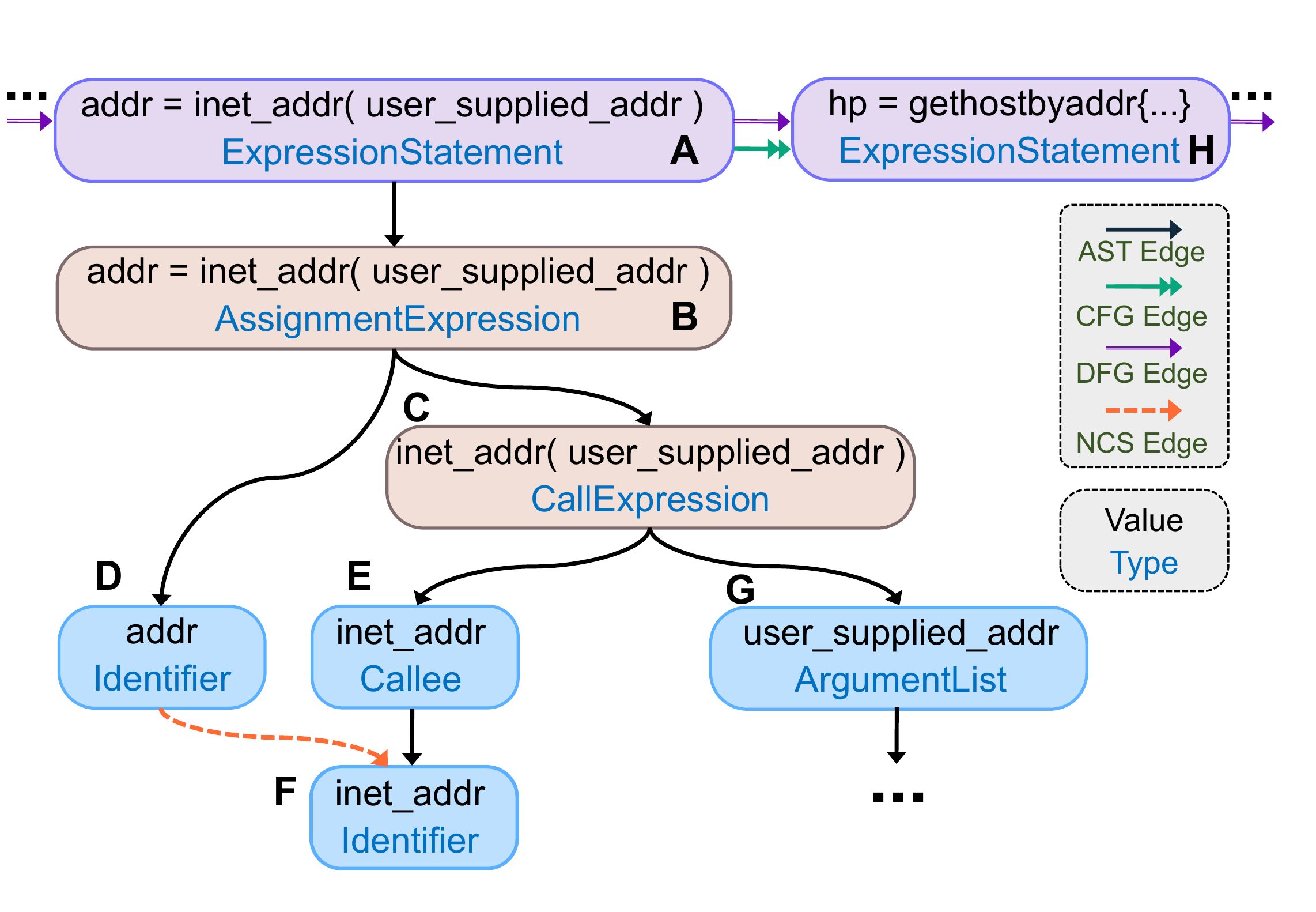}%
\label{motivation_repetitive}}\\

\caption{(a) is a source code snippet of CWE-476.
(b) is a visualisation of the code structure graph of the statements highlighted in red box in (a).
Each node is represented with two attributes: Value (described in the first line) and
Type (descried in the second line). The nodes with different shades
indicate different node types.
}
\label{fig_sim}
\end{figure*}

To accurately detect software vulnerabilities, various vulnerability detection methods based on deep-learning (DL) techniques~\cite{mlmethod1,mlmethod2,mlmethod3}, which aim at learning the patterns of vulnerable code, have been proposed in recent years. They generally process the source code as token sequences~\cite{vulcnn, sysevr, hin2022linevd, fu2022linevul} or code structure graphs~\cite{devign, cao2022mvd, DBLP:journals/tifs/Wang21}. For example,
VulDeePecker~\cite{vuldeepecker} represents
the source code into a sequence of tokens as input and uses a bidirectional Long Short Term Memory (LSTM) model for vulnerability detection. Recent studies~\cite{reveal, IVDETECT,DBLP:journals/corr/empirical} demonstrate that the \textit{structure graph} plays a nonnegligible role in capturing vulnerable code patterns. Reveal~\cite{reveal} leverages code property graph (CPG)~\cite{cpg} by parsing source code and adopts Gated Graph Neural Network (GGNN) to build a vulnerability detection model.
The work Devign~\cite{devign} proposes a joint graph which incorporates four types of edges (\ie Abstract Syntax Tree (AST)~\cite{ast}, Control Flow Graph (CFG)~\cite{cfg}, Data Flow Graph (DFG)~\cite{Dataflow} and Natural Code Sequence (NCS)~\cite{russell}). To learn the structural information, existing studies adopt Graph Neural Networks (GNNs),
such as GGNN, Graph Convolution Network (GCN), \etc achieving state-of-the-art performance in vulnerability detection. These GNNs aggregate nodes~\cite{kipf2016gcn} based on the parent-child relations~\cite{ggnn, gat}, which is beneficial for capturing the adjacent-level information~\cite{Hierarchical}
from the source code.
Despite the promising performance of the existing GNN-based methods, they still have the following limitations:
\textbf{(1) The heterogeneous relations
in the code structure graph are ignored.} 
Previous studies~\cite{devign} generally focus on employing node values in the code structure graph for learning the graph representations. The recent state-of-the-art models have considered the node types~\cite{reveal} or edge types~\cite{wen2022source} in the code structure graph, demonstrating the importance of the structural information for vulnerability detection. However, the studies 
do not jointly consider
different types of nodes and edges, i.e., the heterogeneous relations, which are helpful for capturing the patterns of vulnerable code.
For example, as shown in Fig.~\ref{motivation_repetitive}, we can find that nodes A and B have the same value, but with
different node types
(i.e., $ExpressionStatement$ and $AssignmentExpression$, respectively). Besides, nodes in the graph are connected by different edge types (i.e., AST and CFG, respectively). The heterogeneous relations can enrich the representations of nodes, and thereby are beneficial for venerability detection.
%



\textbf{(2) The long-range dependencies in the graph
are still hard to be captured.}
Most DL-based approaches including the state-of-the art ones~\cite{IVDETECT, reveal} use GNNs~\cite{kipf2016gcn, ggnn} for code vulnerability detection. However, it is well-known that GNNs are limited in handling relationships between distant nodes~\cite{DBLP:conf/www/LiuCZGN21/message, bottleneck}, since GNNs mainly use neighborhood aggregation for message passing. 
Due to a large number of nodes and deep levels in AST-based graphs~\cite{graphcodebert}, the current approaches still face the challenge of learning long-range dependencies in the structure graph by directly adopting GNNs for vulnerability detection~\cite{zhu2021pre,li2018deeper}.





To alleviate the above limitations, in this paper, we present \tool, a \textbf{M}eta-path based \textbf{A}ttentional \textbf{G}raph learning model for code vul\textbf{NE}rability de\textbf{T}ection. Specifically, \tool involves two main components:


\textbf{(1) Multi-granularity meta-path graph construction.}
To exploit the heterogeneous relations in the code structure graph
for vulnerability detection, we design a meta-path graph which jointly involves node types and edge types.
Each meta path in the graph indicates a heterogeneous relation, denoted as a triplet, e.g., ($ExpressionStatement$, AST, $AssignmentExpression$) for the relation between nodes A and B in Fig.~\ref{motivation_repetitive}.
Considering the diversity of node types, \eg there exist 69 node types in the code structure graph of Reveal~\cite{reveal}, the number of heterogeneous relations tends to increase exponentially, which would result in under-fitting for the GNN models~\cite{DBLP:journals/tkde/XiongZKCZ21/underfitting}.
To mitigate the issue, we propose to group the node types into different granularities, including `Statement', `Expression', and `Symbol', thereby reducing the complexity of node types. Specifically, we
construct a multi-granularity meta-path graph for facilitating vulnerability detection.

\textbf{(2) Meta-path based hierarchical attentional graph neural network.}
To learn the representations of the meta-path graph,
we propose a meta-path based hierarchical attentional graph neural network, called \gnn. First, a meta-path attention mechanism is proposed to learn the representation of each meta path, i.e., local dependency, by endowing nodes and edges with different attention weights.
Then, to capture the long-range dependency in the meta-path graph,
we propose a multi-granularity attention mechanism, which captures the importance of heterogeneous relations in different granularities for the final graph representation.

We evaluate \tool on three widely-studied benchmark datasets in software vulnerability detection, including FFMPeg+Qemu~\cite{devign}, Reveal~\cite{reveal}, and Fan \et ~\cite{fan}. We compare with six state-of-the-art software vulnerability detection methods.
The experimental results show that the proposed approach outperforms the state-of-the-art baselines. Specifically, \tool achieves 6.32\%, 21.50\% and 25.40\% improvement comparing with the best baseline regarding the F1 score metric, respectively. In real-world scenarios, \tool detects 27.78\% more vulnerabilities than the best baseline method.

In summary, our major contributions in this paper are as follows:
\begin{enumerate}





\item We propose a novel approach \tool, a meta-path based attentional graph learning model for vulnerability detection. \tool captures heterogeneous relations in the code structure graph by constructing a multi-granularity meta-path graph. 


\item We propose a meta-based hierarchical attentional graph neural network, called \gnn. It can learn the representation of each meta-path and capture the long-range dependency in the meta-path graph.

\item We perform comprehensive experiments for evaluating \tool, which confirms the effectiveness of \tool in code vulnerability detection. We publicly release our code and experimental data for facilitating future research: \textit{{\http}}.

\end{enumerate}

The rest of this paper is organized as follows. Section~\ref{sec:background} describes the background. Section~\ref{sec:architecture} details the two components in the proposed model of \tool,
including the multi-granularity meta-path constructing and meta-path based hierarchical attentional graph neural network. Section~\ref{sec:evaluation} describes the evaluation methods, including the datasets, baselines, implementation and metrics. Section~\ref{sec:experimental_result} presents the experimental results. Section~\ref{sec:discussion} discusses some cases and threats to validity. Section~\ref{sec:conclusion} concludes the paper.


%
%
%
%

\section{Background}
\label{sec:background}




\begin{figure*}[t]
\centering
\includegraphics[width=0.92\textwidth]{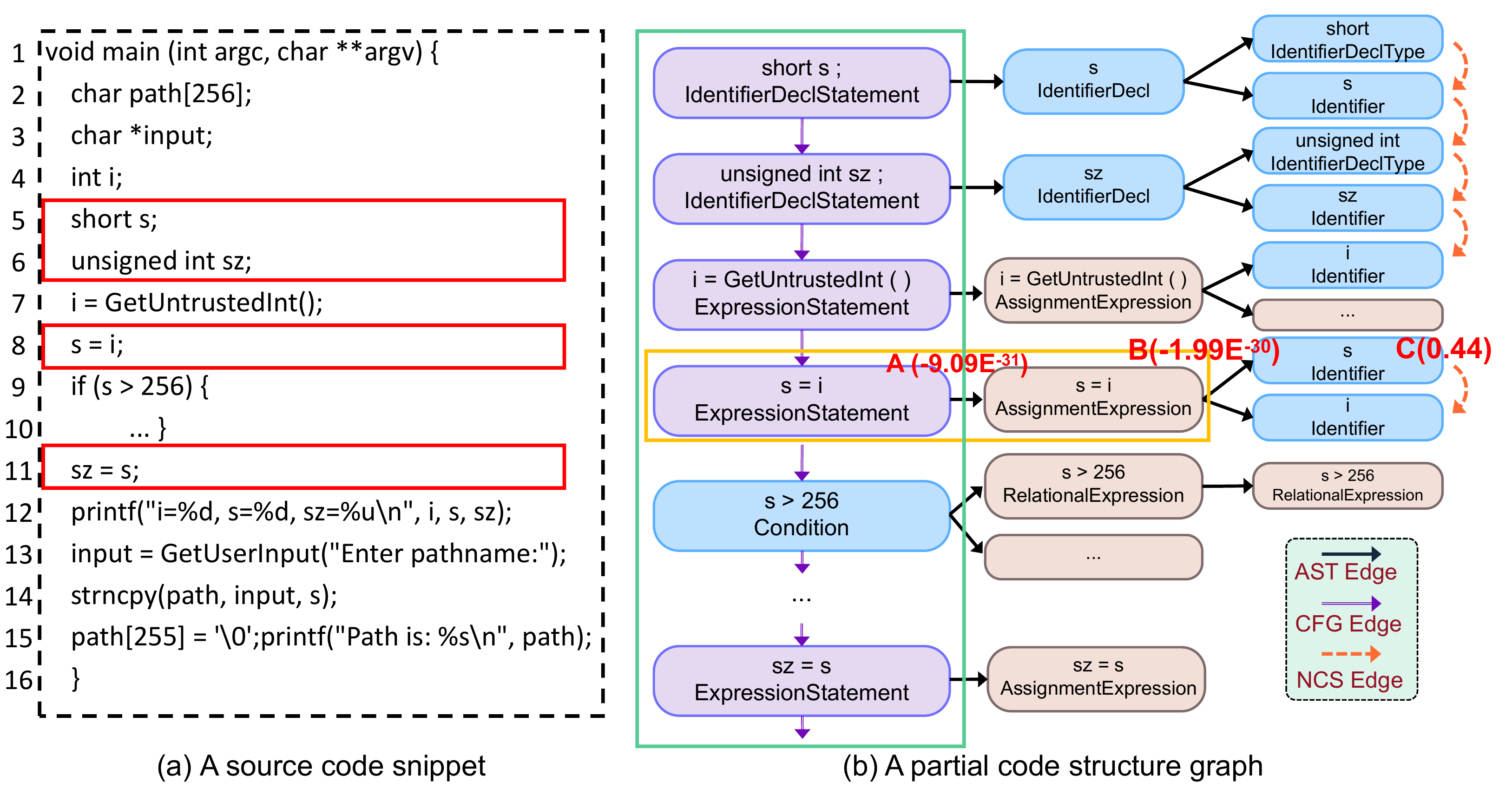}

\caption{(a) is a source code snippet of CWE-839~\cite{CWE-839}.
(b) is the code structure graph generated from the source code snippet.
The red box statement indicates the vulnerable statement. The orange box denotes the same node value and different node types. The green box indicates that GNN needs to use more than six message-passing steps to capture the long dependencies between vulnerability statements.
The red font denotes the meta-path between two nodes, and the brackets indicate the meta-path weight of the path learned in this example.
}

\label{longexample}
\end{figure*}

\subsection{Code Structure Graph}

Code structure graphs are widely used in code vulnerability detection. Devign~\cite{devign} uses code structure graph, which shares the same node set with AST and merges the edge sets of
AST, CFG, DFG, and natural code sequence (NCS). Fig.\ref{motivation_repetitive} shows a code snippet of CWE-476~\cite{CWE-476} and the corresponding code structure graph.
As shown in Fig.\ref{motivation_repetitive}, besides different edge types, each input node is represented with two attributes: \textit{Value} (described in the first line) and \textit{Type} (described in the second line).


The code structure graph encompasses a wealth of both syntactic and semantic information of the source code. 
Following the previous work~\cite{reveal, devign}, we use the Word2Vec~\cite{DBLP:journals/nle/Church17/word2vec} to capture the semantic associations within each node vector.
However, the current DL-based vulnerability detection approaches exhibit limitations in effectively exploiting syntactic information.
For example, IVDetect~\cite{IVDETECT} primarily focuses on combining node types into the node vector, while ignoring the heterogeneous relations, i.e., jointly considering different types of nodes and edges.  Devign~\cite{devign} treats all nodes as the same node type in the Gated Graph Neural Network.



Compared with natural languages, the source code exhibits greater regularity and logical structure~\cite{devign}, meaning that alterations in specific semantic information, such as variable names and identifiers~\cite{DBLP:journals/infsof/RabinBWYJA21/mislead1}, do not necessarily result in vulnerabilities. 
These heterogeneous relations reflect the diverse relationships across various node and edge types, thereby benefiting the acquisition of vulnerability patterns. 
By employing the heterogeneous relations, we can gain deep insights into the variations in node types which are important code structural information~\cite{DBLP:conf/aaai/ZhangLLMLJ20/midslead2}.
In this paper, we aim at exploiting the heterogeneous relations for vulnerability detection by defining meta paths.

\subsection{Graph Neural Networks}
Graph Neural Networks (GNNs) have been widely used in the software engineering tasks, such as code classification~\cite{DBLP:conf/iclr/AllamanisBK18/codec1}, code clone detection~\cite{DBLP:conf/kbse/WuZDYYCLJ20/codeclone1,DBLP:conf/wcre/WangLM0J20/codeclone2,DBLP:journals/tr/HuaSWLX21/codeclone3}, \etc This is due to the inherent ability to
capture the structural information of source code~\cite{reveal}.
GNNs generally aggregate neighbouring nodes' information and use message passing for passing information from the current node to other nodes, finally forming a graph representation.

Several GNN-based methods have been proposed for vulnerability detection.
For instance, Devign~\cite{devign} learns the code structure information by adopting Gated Graph Neural Network (GGNN) to process multiple edge types graphs. Reveal~\cite{reveal} aims at offering a better separability between vulnerable and non-vulnerable samples by using GGNN and multi-layer perceptron (MLP). \ivd~\cite{IVDETECT} uses the feature-attention GCN model
to learn the source code representation, achieving state-of-the-art performance. Despite the good performance, 
the existing methods still struggle to effectively capture long-range dependencies. Previous studies~\cite{DBLP:conf/www/LiuCZGN21/message, bottleneck}  have demonstrated that Graph Neural Networks (GNNs) are limited to learning information from neighboring nodes. In vulnerability detection datasets, the average distance between nodes in different datasets is approximately seven. However, certain traditional GNNs, such as the Graph Convolutional Network, achieve optimal performance with only two layers~\cite{bottleneck}. Consequently, these models can only capture information from nodes with a distance of less than two, which is inadequate for capturing vulnerability patterns. Wang et al.~\cite{Hierarchical} further highlight the challenge of learning long-distance dependencies in code structure graphs, which often have multiple levels and a significant number of nodes.

To illustrate, Fig.~\ref{longexample} shows a portion of the code from line 5 to line 11. In Fig.~\ref{longexample}(a), the three statements highlighted in red boxes (lines 5, 8, and 11) are the root cause of the vulnerability. In the corresponding code structure graph in Fig.~\ref{longexample}(b), the green box indicates that more than six steps of message-passing are required in a GNN to capture the long dependencies between vulnerability statements. Consequently, the model faces difficulties in effectively capturing vulnerability patterns from the code structure graph.

In addition, to evaluate the situation of long-range dependencies, we employ Devign~\cite{devign} as a case study and examine the correlation between model performance and the number of graph nodes. Each node in CFG represents a node, and
the average distance between nodes will grow as
the number of nodes increases~\cite{DBLP:journals/tcas/LuYCC04/small}.
 Our investigation centers on the Reveal dataset \cite{reveal}, and we partition into five intervals based on the number of nodes in code structure graphs, with results shown in Table~\ref{nodenumber}.  Notably, Devign exhibits superior performance for graphs with a lower node count, achieving an accuracy of 90.71\% for graphs containing no more than 50 nodes. However, its performance declines significantly as the number of nodes increases. For graphs with more than 200 nodes, Devign's accuracy drops to approximately 54.17\%. In this paper, we propose a multi-granularity attention mechanism for learning the long-range dependencies.

\begin{table}[]

\centering

\setlength{\tabcolsep}{1.2mm}
\renewcommand{\arraystretch}{1.2}
\caption{The accuracy of Devign~\cite{devign} and \tool on the Reveal
dataset~\cite{reveal} with different numbers of graph nodes.}
\begin{tabular}{l|p{1cm}<{\centering}p{1.cm}<{\centering}p{1.cm}<{\centering}p{1.cm}<{\centering}p{1.cm}<{\centering}}

\toprule
  Node number      & {[}0,50{]} & (50,100{]} & (100,150{]} & (150,200{]} & \textgreater{}200 \\
        \midrule
Devign  &   90.71         &    76.86        &   56.67          &   57.50          &      54.17             \\
\tool & 96.59      & 98.64      & 94.92       & 94.62       & 90.88            \\
\bottomrule
\end{tabular}
\label{nodenumber}
\end{table}


\begin{figure*}[t]
	\centering
    \includegraphics[width=1.0\textwidth]{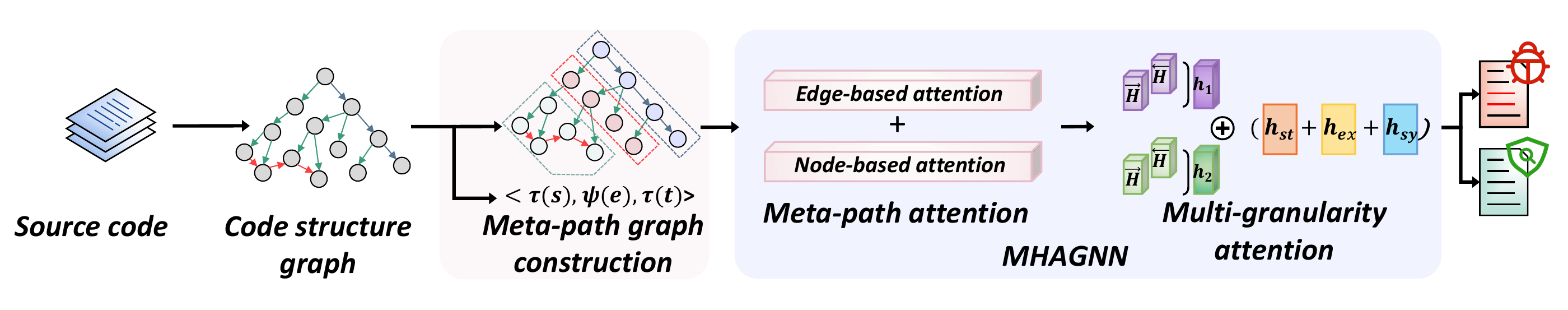}
	\caption{The architecture of \tool.}
	\label{architecture}
\end{figure*}

\begin{table*}[t]
\centering
\setlength{\tabcolsep}{1.1mm}
\renewcommand{\arraystretch}{1.2}

\caption{Classification of node types.}

\begin{tabular}{c|l|c}
\toprule
Node type    &  \multicolumn{1}{c}{Classification}   & Type Number\\
\midrule
Statement & \tabincell{l}{Statement, SwitchStatement, DoStatement, GotoStatement, WhileStatement, BreakStatement, CompoundStatement, \\ForStatement, ReturnStatement,  IdentifierDeclStatement, TryStatement, ClassDefStatement, ContinueStatement, \\IfStatement, ExpressionStatement,  ElseStatement, DeclStatement} & 17\\
\midrule
Expression	 & \tabincell{l}{Expression, InclusiveOrExpression, MultiplicativeExpression, AssignmentExpression, UnaryOperationExpression, \\SizeofExpression,  OrExpression, ShiftExpression, RelationalExpression, CallExpression, CastExpression, \\ConditionalExpression, OperationExpression, EqualityExpression, AdditiveExpression, \\PrimaryExpression, AndExpression, ExclusiveOrExpression, BitAndExpression,  UnaryExpression} & 20\\			
\midrule
Symbol & \tabincell{l}{Symbol, File, IncDec, ForInit, SizeofOperand, PtrMemberAccess, Sizeof, IdentifierDeclType,  IdentifierDecl, ClassDef,
 \\ParameterList, Callee, Condition, ArrayIndexing,  ArgumentList,  Parameter, Argument, ParameterType, CastTarget, \\Function, ReturnType, Label,   FunctionDef, MemberAccess, InitializerList, CFGErrorNode, InfiniteForNode, \\CFGExitNode, CFGEntryNode, Identifier, Decl }& 32\\
\bottomrule
\end{tabular}

\label{nodetype}
\end{table*}

\subsection{Meta-path}

The meta-path has been used in social networks areas~\cite{DBLP:journals/pvldb/SunHYYW11/metapath}, which is a path consisting of a sequence of edge types in the heterogeneous graph. 
The heterogeneous graph contains multiple types of nodes or multiple types of edges with the data structure as a directed graph, where two nodes can be connected via different edges. In vulnerability detection, the code structure graph is also a heterogeneous graph. For example, the previous methods treat the paths AB and AH as the same type path. 
 In the heterogeneous graph, two nodes A and B can be connected via the ``ExpressionStatement-AST-AssignmentStatement'' meta-path in Fig.~\ref{motivation_repetitive}, which is different from the meta-path ``ExpressionStatement-CFG-ExpressionStatement'' connected between nodes A and H. 
Constructing the meta-path in the heterogeneous network can capture the structure information among nodes in the code structure graph. As shown in Figure \ref{longexample}(a), we observe that Line 8 represents a vulnerability statement of source code, as denoted by the orange box containing two nodes in Figure \ref{longexample}(b).
These two nodes possess distinct types even though they share the same node value of ``s=i''. Specifically, the left node corresponds to the ``ExpressionStatement'' type, signifying the statement property, while the right node pertains to the ``AssignmentExpression'' type, representing the expression property. These nodes are connected by an AST edge, forming a heterogeneous relation.  
In this paper, we construct the $<ExpressionStatement, AST, AssignmentExpression>$  meta-path to exploit the structural information in the code structure graph, which captures the heterogeneous relations to facilitate vulnerability detection. 
Without considering the
heterogeneous relations which contain rich structural information within the source code, the previous methods tend to fail for vulnerability detection.
In this paper, we demonstrate the impact of heterogeneous relations on code vulnerability detection. After the model is trained, it adjusts the weights of different heterogeneous relationships, causing the model to focus its attention on specific paths.


\section{Proposed Model}
\label{sec:architecture}

In this section, we introduce the overall architecture of \tool. As shown in Fig.~\ref{architecture}, the architecture includes two main components:
(1) multi-granularity meta-path graph construction, aiming at constructing heterogeneous relations as meta paths.
(2) meta-path based hierarchical
attentional graph neural network, aiming at learning the representations of
the meta-path graph.

\subsection{Multi-granularity Meta-path Graph Construction}

In this section, we first illustrate how to group the node types into multiple granularities, and then describe how we construct the meta-path graph.


\subsubsection{Node Type Grouping}
\label{sec:node}
Directly employing the node types provided by the parsing principles~\cite{cpg} would lead to an increasingly large number of heterogeneous relations. For example, following Reveal~\cite{reveal}, each dataset can be parsed into 69 node types, resulting in more than 10,000\footnote{It contains 69 unique node types and 4 edge types, and the number of types for heterogeneous relations are calculated as $69^{2}*4 = 19044.$} heterogeneous relations. Prior research~\cite{DBLP:journals/tkde/XiongZKCZ21/underfitting} demonstrates that GNN models tend to get underfitting on complex heterogeneous relations. To mitigate the issue, we propose to group the node types into three different granularities, including ``Statement'', ``Expression'' and ``Symbol''.


Specifically, according to the code parsing principles~\cite{cpg}
we group all node types into the following three categories: (1) nodes at ``Statement'' granularity: the node represents the whole sentence in a code snippet, \eg node A and node H in Fig.~\ref{motivation_repetitive}.
(2) nodes at ``Expression'' granularity:
the node consists of two or more operator/operands~\cite{expression}, e.g, the brown-shaded node B and C in Fig.~\ref{motivation_repetitive}.
(3) nodes at ``Symbol'' granularity: the remaining nodes are categorized as ``Symbol'' nodes for simplicity, \eg the nodes D, E, F and G in Fig.~\ref{motivation_repetitive}.
The node types at each granularity are illustrated in Table~\ref{nodetype}, from coarse-grained $Statement$ category to fine-grained $Symbol$ category. The granularity-related categories reflect the structural information of the node value and can facilitate the subsequent DL-based learning process.

\subsubsection{Meta Path Construction}
\label{sec:metapath}
The code structure graph $\mathcal{G}$ is a direct graph, indicated as $\mathcal{G} (\mathcal{V}, \mathcal{E}, \mathcal{A}, \mathcal{R})$, where $\mathcal{V}$, $\mathcal{E}$, $\mathcal{A}$, and $\mathcal{R}$ represent the node set, edge set, node type set, and edge type set, respectively. Each node $v \in \mathcal{V}$ has its associated type with the mapping function $\tau(v): \mathcal{V} \rightarrow \mathcal{A}$. Each edge $e \in \mathcal{E}$ is associated with a type, with the mapping function: $\psi(e): \mathcal{E} \rightarrow \mathcal{R}$.
The edge $e = (s,t)$ denotes the path linked from source node $s \in \mathcal{V}$ to target node $t \in \mathcal{V}$. To capture the structural information of heterogeneous relations between nodes at
different granularity, we propose to build meta paths. Based on the grouped $t_{n}$ node types ($t_{n} = 3$, \ie $Statement$, $Expression$ and $Symbol$) and contained $t_{e}$ edge types ($t_{e} = 4$, \ie $AST$, $CFG$, $DFG$ and $NCS$), we define a meta path as below.


\begin{myDef}[meta path]
     A meta path on the code structure graph is denoted as a triplet $(\tau(s), \psi(e), \tau(t))$, indicating a heterogeneous relation from source node $s$ to a target node $t$ with a connection edge $e$. $\tau(\cdot)$ denotes the type category of the corresponding node, and $\psi(e)$ means the type of the edge $e$.
\end{myDef}

\begin{figure*}[t]
	\centering
    \includegraphics[width=1.0\textwidth]{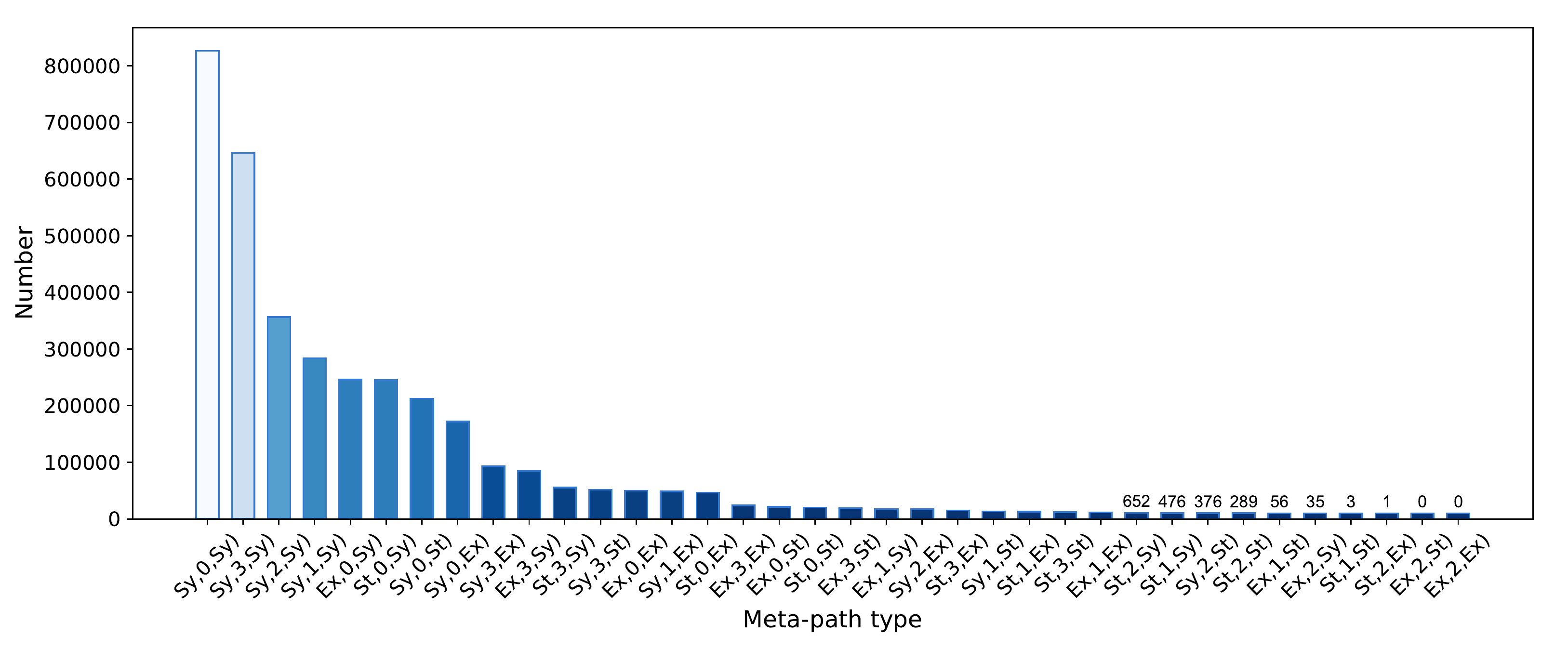}
	\caption{Distribution of different types of meta paths on FFMPeg+Qemu dataset. `$St$' , `$Ex$' and `$Sy$' denote the `$Statement$', `$Expression$' and `$Symbol$' types of nodes. `0', `1', `2', and `3' represent 'AST', `DFG', `CFG' and `NCS' type edges, respectively. The x-axis indicates the type of meta-path, and the y-axis indicates the number of each meta-path. }
	\label{pathnumber}

\end{figure*}

The maximum number of types for the meta paths is $t_{n}^{2} * t_{e} = 36$.
We then analyze the distribution of heterogeneous relations belonging to different meta-path types. Fig.~\ref{pathnumber} illustrate the results on the FFMPeg+Qemu dataset, and the other datasets show a similar distribution trend. As can be seen, the last four types of the total 36 types, \eg $(Ex,2,Ex)$, $(Ex,2,St)$, appear fewer than three times in the dataset. To facilitate the representation learning of the heterogeneous relations in the graph, we filter out the rare types~\cite{DBLP:journals/datamine/WangSLZH18, DBLP:conf/cikm/NingCSHK0HLL022}, and employ the remaining 32 types of meta paths for constructing the meta-path graph.
For the two nodes that have more than one meta path, we keep the multiple meta paths, \eg nodes A and H in Fig.~\ref{motivation_repetitive} have $(St, DFG, St)$ and $(St, CFG, St)$ meta-paths; thus, the structure information of the code strucure graph can be maintained by constructing meta-path graph.

\begin{figure*}[t]
	\centering
    \includegraphics[width=1.0\textwidth]{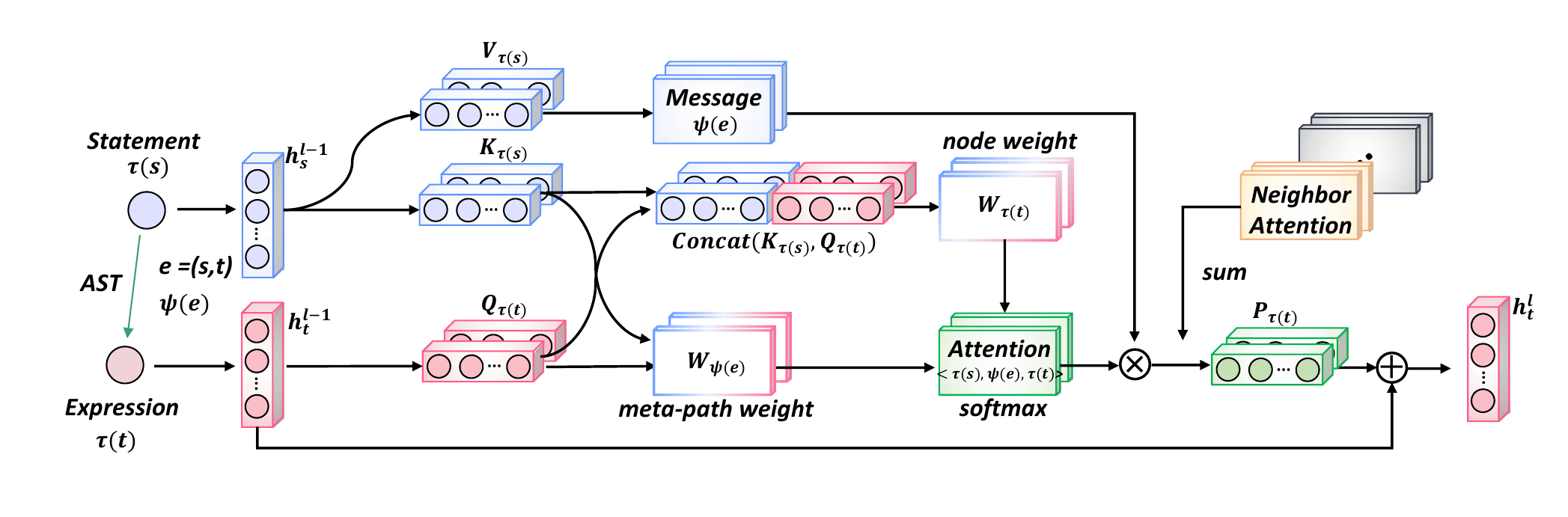}
	\caption{The architecture of \mpa.}
	\label{mpa}
\end{figure*}

\subsection{Meta-path based Hierarchical Attentional Graph Neural Network}

In this section, we illustrate the proposed meta-path based hierarchical attentional graph neural network, named \gnn. 
\gnn consists of two modules: (1) a meta-path attention mechanism for capturing the representations of heterogeneous relations; and (2) a multi-granularity attention mechanism for capturing long-range dependency in the meta-path graph.

\subsubsection{Meta-path Attention Mechanism}
To better utilize the heterogeneous relations, i.e., triplets $(\tau(s), \psi(e), \tau(t))$, from the constructed multi-granularity meta-paths
, we devise a meta-path attention mechanism, with detailed architecture presented in Fig.~\ref{mpa}.
 The \mpa consists of node-based attention and edge-based attention, which aims at learning the importance of different node types and different edge types in the graph structure representation. 
 
\textbf{Node-based attention:}
The node-based attention score $Att_{node}^{l}$ for the target node $t$ in the $l$-th layer is defined as follows:
\begin{equation}
\label{node-att}
Att_{node}^{l}=\sigma \left(W^l_{\tau(t)} \cdot \left ( {K^l}(s)|| {Q^l}(t) \right) \right) 
\end{equation}
\begin{equation}
\label{node-K}
K^l{(s)} = Linear_{\tau(s)}(h_{s}^{l-1})
\end{equation}
\begin{equation}
\label{node-Q}
Q^l{(t)} = Linear_{\tau(t)}(h_{t}^{l-1})
\end{equation}
where $W^l_{\tau(t)}$ is a trainable weight matrix, indicating the contribution of node type $\tau(t)$ to the representation of the whole graph. The symbol $||$ is the concatenation operation and $\sigma$ is the sigmoid activation function. $Q^l{(t)}$ and $K^l{(s)}$ are the linear projection of node vector $h^{l-1}_t$ and $h^{l-1}_s$, respectively, which are used to learn the textual information of the code.
$Q^l{(t)}$ and $K^l{(s)}$ are the linear projection of node $t$ and $s$, respectively. 
In Equation~(\ref{node-Q}) and~(\ref{node-K}), $Linear$ denotes a fully connected neural network layer. $h_{t}^{l-1}$ and $h_{s}^{l-1}$ denote the node vectors of $t$ and $s$ in $(l-1)$-th layer, respectively, where $h_t^{0}$ and $h_s^{0}$ are initialized as 100-dimensional vector by word2vec~\cite{DBLP:journals/nle/Church17/word2vec}.


\textbf{Edge-based attention:} The edge-based attention score $Att_{edge}^{l}$ for the target node $t$ is defined as following:

\begin{equation}
\label{edge-att}
Att_{edge}^{l}=\left( {K^l}(s){W_{\psi(e)}}{Q^l}(t)^{T}\right) \cdot \frac{\mu (\psi(e))}{\sqrt{\frac{d}{h}}}
\end{equation}
where $W_{\psi(e)}$ and $\mu(\psi(e))$ denote the trainable matrix and parameter for each edge type $\psi(e)$, representing the importance of the edge type to the source node $s$ and the target node $t$, respectively.
$d$ and $h$ denote the vector dimension and the number of edge-based attention heads, respectively. 

\textbf{Meta-path attention:} Based on the computed node-based attention and edge-based attention, we devise the meta-path attention score $Att_{(\tau(s),\psi(e),\tau(t))}^l$ for the target node $t$ to model the heterogeneity of the relationship $(\tau(s), \psi(e), \tau(t))$:
\begin{equation}
\label{meta-att}
Att_{(\tau(s),\psi(e),\tau(t))}^l = softmax\left (\|_{1}^{H}\left(Att_{edge}^{l} + 
Att_{node}^{l} \right)\right)
\end{equation}
where $||_{1}^{H}$ represents the concatenation of
$H$ attention heads.

Finally, we sum the attention scores of all neighbor nodes connected to the node $t$, which is used as the meta-path attention score of the node $t$.
We treat meta-path attention score $Att_{(\tau(s),\psi(e),\tau(t))}^l$ as input and utilize message passing from source nodes to the target nodes, which incorporates the heterogeneous relations into $l$-th layer's.
To enhance the ability of GNN to represent different nodes, we also establish a residual connection~\cite{DBLP:conf/eccv/HeZRS16/res} with the previous $(l-1)$-th layer's output, and get the updated node vector $h^{l}_t$ as:

\begin{equation}
\begin{aligned}
\label{ht}
h_t^{l} = \sigma{ \left ( \sum_{s \in N_{t}} Att_{ (\tau (s), \psi (e),\tau (t) ) }^{l} \cdot {V^l}(s) 
\right )  } 
+ h_{t}^{(l-1)}
\end{aligned}
\end{equation}
\begin{equation}
V^l{(s)} = Linear_{\tau(s)}(h_{s}^{l-1})
\end{equation}
where $N_{t}$ denotes the set of neighboring nodes of node $t$ and $h_{s}^{l-1}$ denotes the vector representation of the node $s$
in $(l-1)$-th layer. $V^l{(s)}$ is also the linear projection of node $s$. 

\subsubsection{Multi-granularity Attention}
To enhance the representations of nodes at different granularities, we propose to adopt both the average-pooling layer and the max-pooling layer simultaneously. The average-pooling layer~\cite{DBLP:journals/tcsv/SunCWZHY21/avepool} is designed to capture the long-range dependency in the meta-path graph; while the max-pooling layer aims at magnifying the contribution of the key nodes considering that only few nodes in the graph are vulnerability-related~\cite{DBLP:conf/eccv/WooPLK18/cbam}. The multi-granularity attention score $M$ is calculated as below:
\begin{equation}
\begin{aligned}
\label{cbam}
M = \sigma \left ( MLP\left (\omega_{1,i} \cdot AvgPool(F_{i}) + \omega _{2,i} \cdot MaxPool(F_{i})\right ) \right ) 
\\(i = st, ex, sy)
\end{aligned}
\end{equation}

\noindent where $\omega$ 
denotes a trainable weight for different levels average-pooled and max-pooled features and $i$ denotes different levels (\ie granularities) of nodes. $F_{st}$, $F_{ex}$, $F_{sy}$ are the node type representation of ``Statement'', ``Expression'' and ``Symbol'', respectively, which are calculated as $F_{i} = \left \{ {h_{t}}\right\}_{q=1}  ^{\left|\mathcal{V}_{i} \right|}, (i = st, ex, sy)$. 
$\left|\mathcal{V}_{i}\right|$ denotes the number of a single node types in the whole graph.
 The single node vector $h_t$ is concatenated by bidirectional GRU~\cite{DBLP:journals/corr/BahdanauCB14/gru1}, calculated as $h_t = \left(\overrightarrow{GRU}\left(h_t^l \right) \right)||\left({ \overleftarrow{GRU}\left(h_t^l\right)}\right)$.
Finally, we use the typical
CrossEntropy loss function~\cite{DBLP:journals/anor/BoerKMR05/ce} for vulnerability prediction.



\section{Evaluation}
\label{sec:evaluation}
\subsection{Research Questions}
We evaluate the \tool with the state-of-the-art vulnerability methods and aim at answering the following questions:

\begin{enumerate}[label=\bfseries RQ\arabic*:,leftmargin=.5in]
    \item How does our method \tool perform in vulnerability detection?
    \item What is the impact of different modules
    in \gnn
    on the detection performance of \tool?
    \item How effective is \tool for detecting Top-25 Most Dangerous CWE?
    \item What is the influence of hyper-parameters on the performance of \tool?
\end{enumerate}

\subsection{Experiment Setup}
\subsubsection{Dataset}

\begin{table}[t]
\centering
\setlength{\tabcolsep}{1.1mm}
\renewcommand{\arraystretch}{1.2}

\caption{Statistics of the datasets.}
\begin{tabular}{c|r|c|c}
\toprule
Dataset & \multicolumn{1}{c|}{Samples} &  \multicolumn{1}{c|}{Ratio (\#Vul:\#Non-vul)} & \multicolumn{1}{c}{Language}\\

\midrule
FFMPeg+Qemu~\cite{devign}  & 22,361    & 1:1.2 & C
\\
Reveal~\cite{reveal}  & 18,169   & 1:9.9  & C/C++       \\
Fan \et~\cite{fan}     & 179,299   & 1:16  & C/C++       \\
\bottomrule

\end{tabular}
\label{dataset}
\end{table}

\begin{table*}[h]
\centering

\setlength{\tabcolsep}{1.2mm}
\renewcommand{\arraystretch}{1.2}

\caption{
Comparison results between \tool and the baselines on the three datasets. ``-'' means that the baseline does not apply to the dataset in this scenario. The best result for each metric is highlighted in bold. The shaded cells represent the performance of the top-3 best methods in each metric. Darker cells represent better performance.}
\resizebox{.97\textwidth}{!}{
\begin{tabular}{l|cccc|cccc|cccc}
\toprule
\diagbox{Metrics(\%) }{Dataset} & \multicolumn{4}{c|}{FFMPeg+Qemu \cite{devign}}        & \multicolumn{4}{c|}{Reveal \cite{reveal}}             & \multicolumn{4}{c}{Fan \et. \cite{fan}}                \\
\midrule
Baseline                         & Accuracy & Precision & Recall & F1 score     & Accuracy & Precision & Recall & F1 score    & Accuracy & Precision & Recall & F1 score    \\
\midrule
VulDeePecker                    & 49.61   & 46.05    & 32.55 & 38.14 & 76.37   & 21.13    & 13.10 & 16.17 & 81.19   & \cellcolor{gray!70}\textbf{38.44}    & 12.75 & 19.15 \\

Russell \et                  & \cellcolor{gray!20}57.60   & \cellcolor{gray!20}54.76    & 40.72 & 46.71 & 68.51   & 16.21    & \cellcolor{gray!20}52.68 & 24.79 & 86.85   & 14.86    & \cellcolor{gray!20}{26.97} & 19.17 \\

SySeVR                          & 47.85   & 46.06    & 58.81 & 51.66 & 74.33   & \cellcolor{gray!45}40.07  & 24.94 & 30.74   & \cellcolor{gray!20}90.10   & \cellcolor{gray!45}30.91   & 14.08 & 19.34 \\

Devign                          & 56.89   & 52.50    &  \cellcolor{gray!20}64.67 &  \cellcolor{gray!20}57.95 & \cellcolor{gray!45}87.49   & \cellcolor{gray!20}31.55    & 36.65 & \cellcolor{gray!20}33.91 & \cellcolor{gray!70}\textbf{92.78}   & \cellcolor{gray!20}30.61   & 15.96   & \cellcolor{gray!20}20.98 \\

Reveal                          & \cellcolor{gray!45}61.07   & \cellcolor{gray!45} 55.50    & \cellcolor{gray!45} 70.70 & \cellcolor{gray!45}62.19 &  \cellcolor{gray!20}81.77   &  \cellcolor{gray!20}31.55    & \cellcolor{gray!45}{61.14}& \cellcolor{gray!45}41.62   & 87.14   & 17.22   & \cellcolor{gray!45}34.04 & \cellcolor{gray!45}22.87 \\

IVDetect                        & 57.26   & 52.37    & 57.55 & 54.84 & -   & -         & -      & -      & -   & -         & -      & -      \\
\midrule
\tool
& \textbf{\cellcolor{gray!70}63.28}   & \cellcolor{gray!70}\textbf{56.27}    & \cellcolor{gray!70}\textbf{80.15} &\cellcolor{gray!70} \textbf{66.12} &\cellcolor{gray!70}\textbf{91.60}   & \cellcolor{gray!70} \textbf{42.86}    &  \cellcolor{gray!70}\textbf{61.68} & \cellcolor{gray!70}\textbf{50.57} & \cellcolor{gray!45}{91.38}   & 22.71  & \cellcolor{gray!70}\textbf{38.92} & \cellcolor{gray!70}\textbf{28.68}\\
\bottomrule
\end{tabular}}

\label{e1_effective}
\end{table*}

In the experiments, we conduct \tool on three vulnerability datasets, \ie FFMPeg+Qemu~\cite{devign}, Reveal~\cite{reveal}, and Fan \et~\cite{fan}.  
The statistics
of the three datasets are listed in detail in Table~\ref{dataset}.
The FFMPeg+Qemu dataset was proposed by
Devign~\cite{devign},  which contains 10+k vulnerable and 12+k non-vulnerable entries.  It is a relatively balanced dataset and 45.0\% of the samples are vulnerable. The Reveal and Fan \et datasets are imbalanced, which contain +18k samples with 9.16\% vulnerable and +179k samples with 5.88\% vulnerable samples. 
For the programming languages, FFMPeg+Qemu only comes from open-source C projects, and the others datasets collect open-source C/C++ projects.

\subsubsection{Baseline Methods}
We consider the token-based methods and graph-based methods in vulnerability detection. We implement the baselines and their corresponding parameter settings based on the original papers as much as possible. Since Devign~\cite{devign} did not publish their source code and parameter settings, we used the repository reproduced by Reveal~\cite{reveal}.

\textbf{Token-based methods}: \textbf{VulDeePecker}~\cite{vuldeepecker} splits source code into program slices with control dependencies incorporated. 
The program slices are fed into the model built with
an LSTM and attention mechanism. \textbf{Russell \et}~\cite{russell} embed the labeled source code into the corresponding matrix. They detect code vulnerabilities by CNN, integrated learning, and random forest classifiers. \textbf{SySeVR} uses multiple code features as input (\ie code statements, program dependencies, and program slices) and utilizes a bidirectional RNN for code vulnerability detection.

\textbf{Graph-based methods}: \textbf{Devign}~\cite{devign} uses the GGNN approach on the code structure graph for vulnerability detection. \textbf{Reveal}~\cite{reveal} uses Code Structure Graph (CPG) as the input and leverages GGNN in the feature extraction step. And it combines MLP and Triplet Loss during the training phase. \textbf{\ivd}~\cite{IVDETECT} constructs the Program Dependency Graph (PDG) 
and uses GCN to learn the graph representation for capturing vulnerable pattern.

\subsubsection{Implementation}
We implement our model \tool in Python 3.7 using PyTorch~\cite{DBLP:conf/nips/PaszkeGMLBCKLGA19/pytorch} and Deep Graph Library (DGL)~\cite{DBLP:journals/corr/abs-1909-01315/dgl}. We train our model with the NVIDIA GeForce RTX 3090 GPU, installed with Ubuntu 20.04 and CUDA 11.4. In the embedding layer, the initial input dimension $d$ is set to 100 and the hidden state dimension is set to 64. The number of \gnn layers is set to 2 and the head of meta-path attention is set to $h$ = 4. We adopt Adam optimizer~\cite{DBLP:conf/iwqos/Zhang18/adam} to train our model with a learning rate $5e^{-4}$. The batch sizes for FFMPeg+Qemu, Reveal, and Fan \et datasets are set as 512, 512 and 256, respectively.

In addition, to ensure the fairness of the experiments, we use the same data splitting for all baseline approaches as \tool. We randomly partition the dataset into disjoint train, valid, and test sets with the ratio of 8:1:1.

\subsection{Evaluation Metrics}

We use the following four widely-used evaluation metrics to measure our model's performance:

\textbf{Precision:} Precision is the percentage of true vulnerabilities among the vulnerabilities retrieved. $TP$ and $FP$ are the number of true positives and false positives, respectively.
\begin{equation}
\label{precision}
Precision = \frac{TP}{TP+FP}
\end{equation}

\textbf{Recall:} Recall is the percentage of vulnerabilities that are retrieved out of all vulnerable samples. $TP$ and $FN$ are the number of true positives and false negatives, respectively.

\begin{equation}
\label{recall}
Recall = \frac{TP}{TP+FN}
\end{equation}

\textbf{F1 score:} F1 score is the harmonic mean of precision and recall metrics.

\begin{equation}
\label{f1}
F1\ score = 2 \times \frac{Precision\times  Recall}{Precision+Recall}
\end{equation}

\textbf{Accuracy:} Accuracy is the percentage of correctly classified samples to all samples.
$TN$ is the number of true negatives and $TP+TN+FN+FP$ represents the all samples.

\begin{equation}
\label{Acc}
Accuracy = \frac{TP+TN}{TP+TN+FN+FP}
\end{equation}

\section{Experimental Result}
\label{sec:experimental_result}
 \subsection{RQ1: How does our method \tool perform in vulnerability detection?}
To answer this research question, we first explore the performance of \tool and compare it with other baseline methods. Then, we visualize the features learned by the \tool to verify the validity of the learned vulnerability patterns.
 
 \subsubsection{Effectiveness of \tool}
 
Table~\ref{e1_effective} shows the overall results of all baseline models and \tool on the four
evaluation metrics.
Overall, \tool achieves better results and outperforms all of the six referred token-based and graph-based approaches on FFMPeg+Qemu, Reveal and Fan \et dataset in terms of F1 score by 6.32\%, 21.50\% and 25.40\%, respectively. 
For the four performance metrics on the three datasets, \tool{} has the best performance in 10 out of the 12 cases. Compared with the best-performing baseline Reveal, our method obtains an average performance improvement of 6.84\%, 23.04\%, 9.53\% and 17.74\% on the four metrics, respectively.

We observe that \tool outperforms all the baseline methods on the three datasets in terms of F1 score and recall metric. Compared with the best-performing baseline method, \tool achieves an average absolute improvement of 6.23\% with respect to the F1 score on the three datasets. \tool also improves the recall metric by 13.37\%, 0.88\% and 14.34\%, respectively, over the best baseline methods. 
Such improvement benefits the 
scenario of vulnerability detection, since a higher recall indicates a larger percentage of vulnerabilities that can be detected.

MAGNET achieves the best performance on all metrics for the FFMPeg+Qemu and Reveal datasets. However, on the Fan et al. dataset, it performs worse than the IVDetect. We believe that this may be due to the class imbalance problem~\cite{reveal} of the Fan et al dataset. MAGNET focuses more on vulnerabilities, which makes it higher on the Recall metric. In addition, in Table~\ref{e1_effective}, the symbol ``-'' indicates that IVDetect could not converge on the Reveal and Fan et al. datasets, and would identify all code snippets as non-vulnerable when they use the same split setting with \tool. The reason may be attributed to that IVDetect is designed for balanced training datasets in the original paper~\cite{IVDETECT}, and tends to fail for the commonly-used imbalanced datasets such as the Reveal and Fan et al. datasets. 

Experimental results also show that three graph-based methods Devign, Reveal and IVDetect outperform the three token-based methods. For example, the gray cells represent the top-3 best results
in the FFMPeg+Qemu dataset
appear 6 times in the graph-based methods but only twice
in the token-based methods. 
The reason may be attributed to that token-based methods are more advantageous in capturing the sequence information of codes while ignoring the structural information. In contrast, graph-based methods can better capture code structure information, which is beneficial for vulnerability detection.


 \subsubsection{Result Visualization}
To further analyze the effectiveness of \tool,
 we visualize the representations learnt by \tool via the popular t-SNE technique~\cite{van2008visualizing/TSNE}. For comparison, we also use t-SNE to visualize existing graph-based code vulnerability detection methods as well. In the t-SNE space, a larger distance between different classes (\ie vulnerable and non-vulnerable examples) of nodes indicates a clear and greater separability of classes,
 which leads to a higher performance of vulnerability detection.
 In addition, to facilitate the quantification, we use centroids distance $D$~\cite{DBLP:conf/nips/MaoZYVR19/distance} for quantitatively illustrating
 the separability between different classes. 
 
\begin{figure*}[t]
\centering
\subfloat[Devign $D= 0.0108$]{\includegraphics[width=0.48\linewidth]{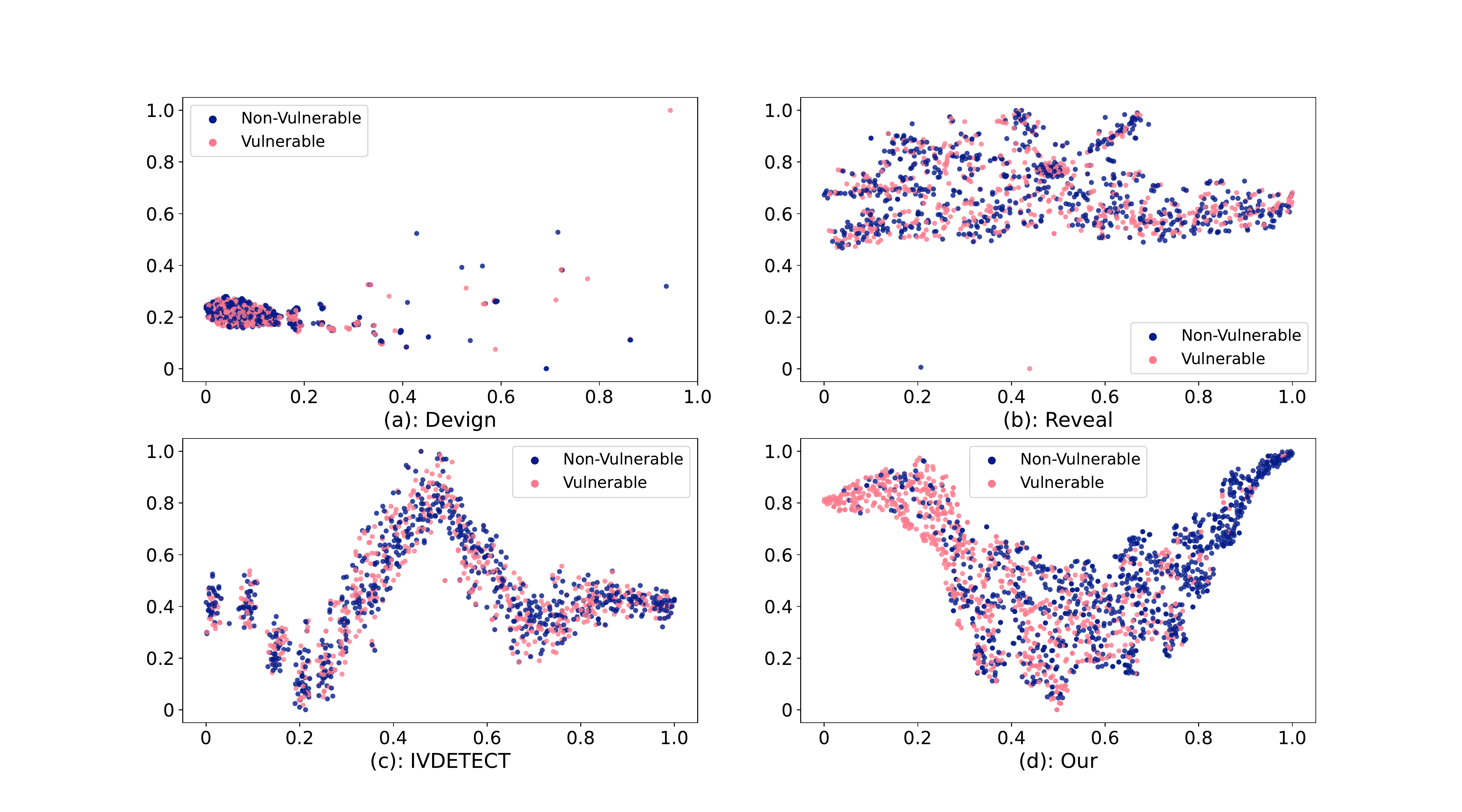}%
\label{fig_first_case}}
\hfil
\subfloat[Reveal $D= 0.0427$]{\includegraphics[width=0.47\linewidth]{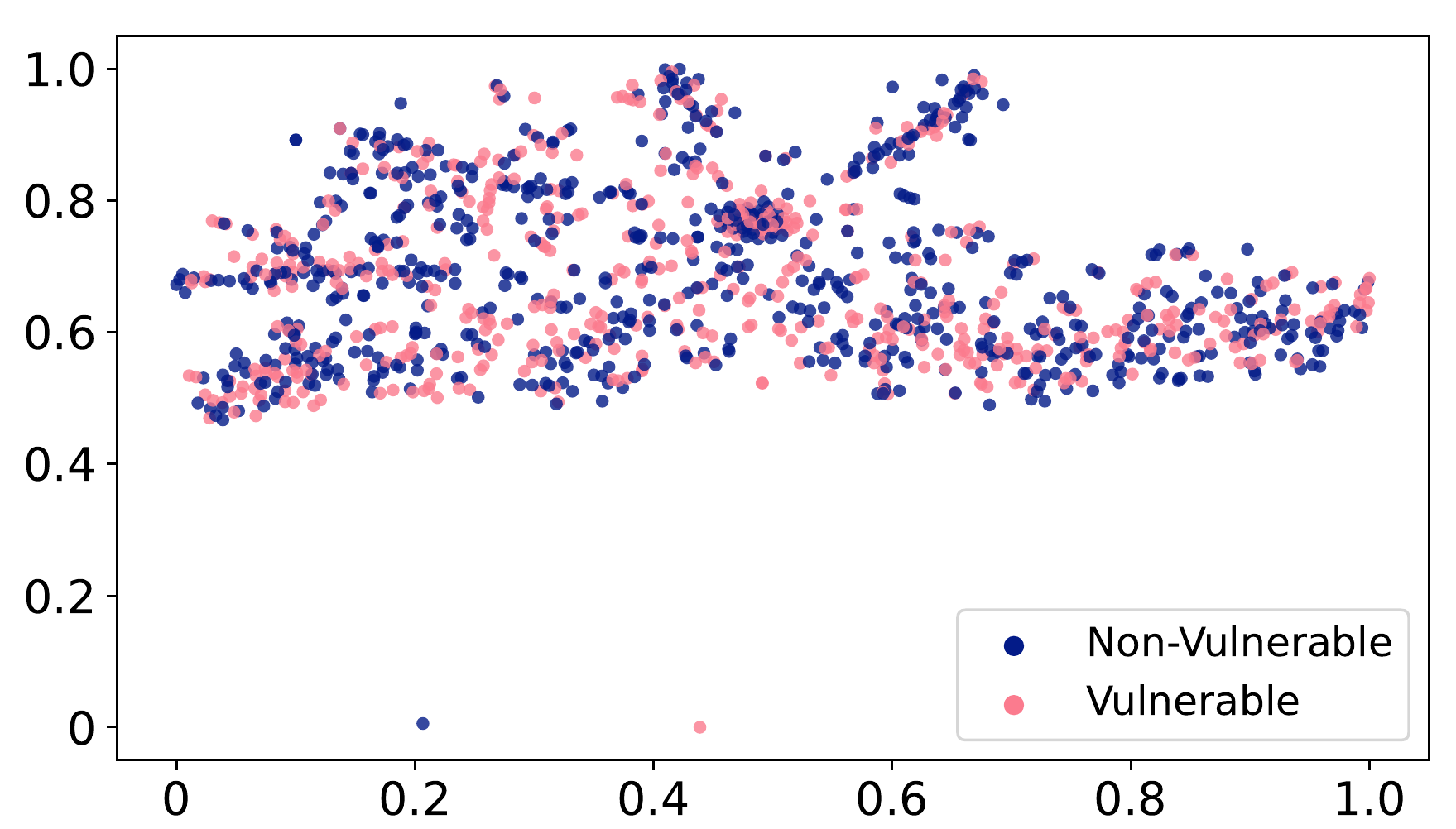}%
\label{fig_second_case}}\\
\subfloat[\ivd $D = 0.0110$]{\includegraphics[width=0.48\linewidth]{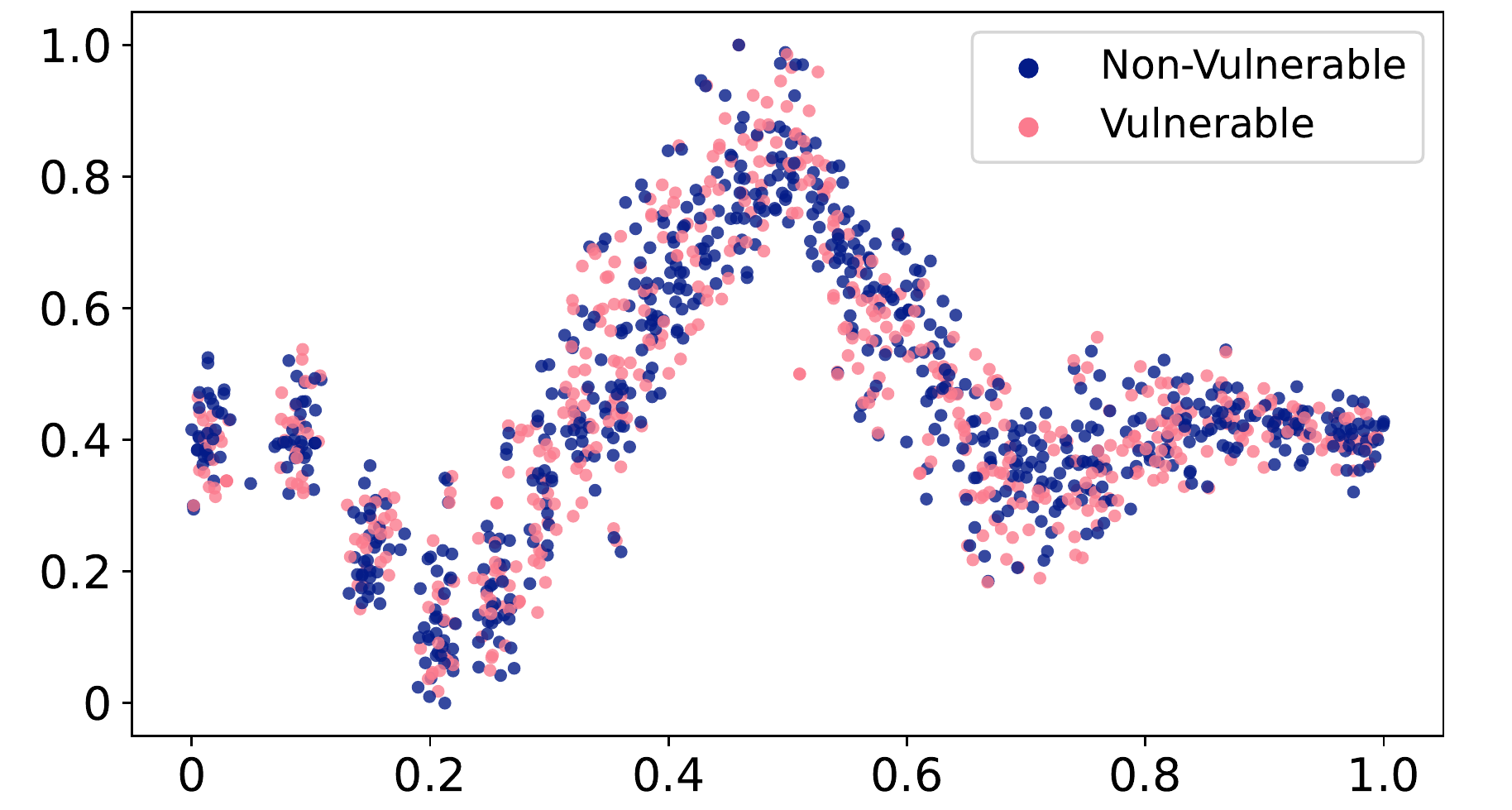}%
\label{fig_third_case}}
\hfil
\subfloat[\tool $D= 0.2901$]{\includegraphics[width=0.47\linewidth]{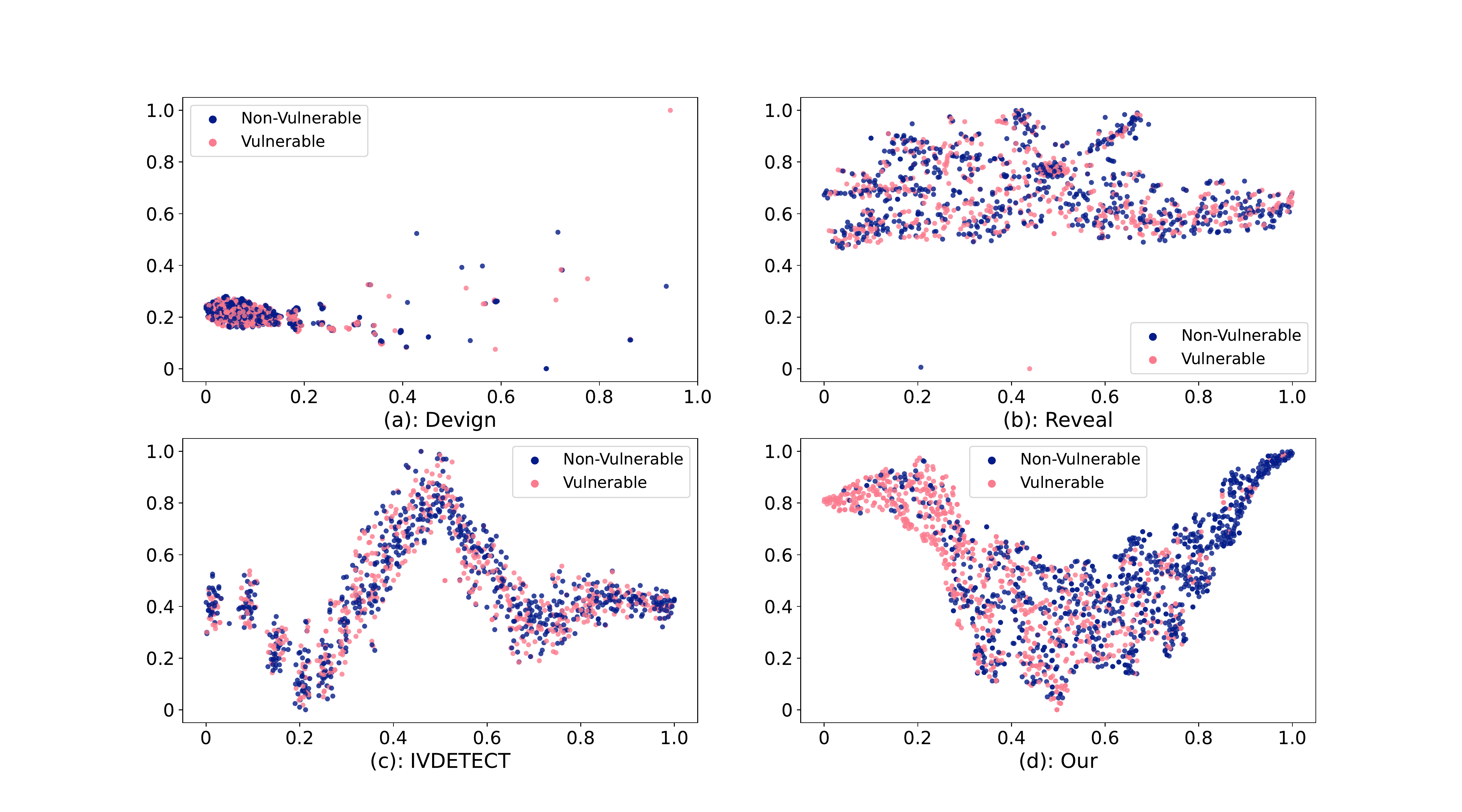}%
\label{fig_my_case}}
\caption{The t-SNE~\cite{van2008visualizing/TSNE} plot illustrates the distribution between vulnerable (pink) and non-vulnerable (dark blue) examples in the code representations of the different approaches. $D$ indicates the centroids distance between the centers of the vulnerable and non-vulnerable examples.}
\label{fig_sim}
\end{figure*}

 The feature visualization graphs of t-SNE for the existing graph-based models are shown in Fig.~\ref{fig_first_case} to~\ref{fig_third_case}. It shows that the positive and negative samples in Devign are thoroughly mixed, with the central distance at only 0.0108. Compared with Devign, both Reveal and \ivd obtain larger central distances, and the scatter appears more dispersed but still lacks
 the separability visible to the naked eye. In Fig.~\ref{fig_my_case}, we show the separability of our \tool. We can observe that the left side aggregates more vulnerability samples, while the right side has more non-vulnerable examples. Besides, \tool shows the largest central distance at 0.2901 among all the methods. The visualization further demonstrates the effectiveness of \tool in distinguishing vulnerable code from non-vulnerable code.

 \begin{tcolorbox}
 \textbf{Answer to RQ1:} \tool outperforms all baseline methods in terms of recall and F1-score. On the F1 score, \tool improved by 6.32\%, 21.50\% and 25.40\% compared with the best baseline Reveal on the three datasets, respectively. 
 Visualization further demonstrates that
 \tool distinguishes code vulnerabilities better than the
 baselines.
 \end{tcolorbox}

 \subsection{RQ2: What is the impact of different modules of \gnn?}
 
To answer this research question, we 
explore the effect of each module in \gnn on the performance of \tool by performing ablation study on all three datasets.

\begin{table*}[t]
\centering

\setlength{\tabcolsep}{1.2mm}
\renewcommand{\arraystretch}{1.2}

\caption{Results of ablation study.}

\resizebox{.97\textwidth}{!}{
\begin{tabular}{l|cccc|cccc|cccc}
\toprule
\diagbox{Metrics(\%) }{Dataset} & \multicolumn{4}{c|}{FFMPeg+Qemu \cite{devign}}        & \multicolumn{4}{c|}{Reveal \cite{reveal}}             & \multicolumn{4}{c}{Fan \et. \cite{fan}}                \\
\midrule
Metrics      & Accuracy      & Precision      & Recall     & F1        & Accuracy     & Precision    & Recall    & F1       & Accuracy    & Precision    & Recall   & F1      \\
\midrule
w/o edge-att & 61.16         & 55.96          & 69.52      & 62.01     & 91.02        & 39.60        & 55.14     & 46.09    & 90.54       & 19.57        & 36.15    & 25.39   \\
w/o node-att & 60.97         & 55.97          & 62.70      & 59.15     & 90.89        & 39.62        & 58.88     & 47.37    & 91.27       & 19.47        & 30.62    & 23.80   \\
w/o multi-att    & 60.94         & 55.77          & 60.92      & 58.23     & 88.41        & 31.22        & 55.14     & 39.86    & 86.27       & 12.92        & 36.31    & 19.06   \\
\midrule
\tool & \textbf{63.28}         & \textbf{56.27}          & \textbf{80.15}      & \textbf{66.12}     & \textbf{91.60}        & \textbf{42.86}        & \textbf{61.68}     & \textbf{50.57}    & \textbf{91.38}       & \textbf{22.71}        & \textbf{38.92}    & \textbf{28.68}  \\
\bottomrule
\end{tabular}}

\label{e2_ablation}
\end{table*}

We construct the following three variations of \tool for comparison:
(1) without edge-based attention (denoted as
\textit{w/o edge-att}): 
we remove the edge-based attention to validate the impact
of the edge-based attention;
(2) without node-based attention (denoted as \textit{w/o node-att}): 
we remove the node-based attention to verify the impact
of node-based attention;
(3) without the multi-granularity attention (denoted as \textit{w/o multi-att}): 
we obtain the graph representation through simply summing the node feature embeddings instead of using the proposed multi-granularity attention.
 
The results of the different variants are
shown in Table~\ref{e2_ablation}. We find that the performance of all the variants is lower than that
of \tool, which indicates that all the modules contribute to
the overall performance of \tool. Specifically, without the edge-based attention, the results of accuracy, precision, recall and F1 score on the three datasets drop by 1.18\%,  2.24\%, 6.65\%, and 3.57\% on average, respectively. Without node-based attention, the four metrics decrease by 1.04\%, 2.26\%, 9.52\%, and 5.02\%, respectively. The node-based attention and edge-based attention contribute greatly to the model performance, since they capture different types of structural information in the meta-path graph. 

Among all the three parts, the multi-granularity attention layer contributes the most to the overall performance, which improves the F1 score by 11.93\%, 21.18\%, and 33.54\% on the three datasets, respectively. The reason may be attributed to that the multi-granularity attention facilitates the learning process of global long-range dependency, and can better capture the patterns of vulnerable code.

 \begin{tcolorbox}
 \textbf{Answer to RQ2:} 
The various components of the \gnn effectively improve the \tool performance. The multi-granularity attention layer contributes the most to the overall performance.
 \end{tcolorbox}
 
 \subsection{RQ3: How effective is \tool for Top-25 Most Dangerous CWE?}
 
\begin{table}[t]
\centering
\setlength{\tabcolsep}{1.2mm}
\renewcommand{\arraystretch}{1.2}
\caption{The accuracy of baseline and \tool for the top-25 most dangerous CWEs. Due to the small number of partial vulnerabilities, we only show the types of vulnerabilities with more than 50 samples. Percentage indicates the proportion of vulnerabilities in the samples.}
\scalebox{0.95}{\begin{tabular}{l|r|c|c|c|c}

\toprule
 CWE Type & Percentage & Devign  & Reveal  & \ivd & \tool \\
\midrule
 CWE-787 &	3.34\% & 45.37 & 51.71 & 56.60             & \textbf{83.19}              \\
 CWE-79 &	1.16\%  & 39.13 & 65.22 & 51.43             & \textbf{71.79}              \\
 CWE-125&	8.61\%  & 47.14 & 52.54 & 61.08             & \textbf{72.86}              \\
 CWE-20 &	23.48\%  & 42.00 & 53.58 & 57.88             & \textbf{75.32}              \\
 CWE-416 &	11.78\%  & 46.67 & 58.13 & 61.05             & \textbf{72.97}              \\
 CWE-22 &	0.77\%   & 56.86 & 47.06 & 48.00             & \textbf{64.00}              \\
 CWE-190 &	3.84\% & 44.11 & 52.47 & 59.20             & \textbf{76.34}              \\
 CWE-287 &	0.72\% & 35.14 & 48.65 & 51.28             & \textbf{71.43}              \\
 CWE-476 &	5.33\% & 48.00 & 49.41 & 63.22             & \textbf{73.54}              \\
 CWE-119&	28.61\%
  & 43.77 & 51.23 & 60.83             & \textbf{77.07}              \\
 CWE-200&	9.02\%
  & 44.59 & 51.18 & 56.17             & \textbf{76.82}              \\
 CWE-732 &	1.64\%
 & 34.38 & 56.25 & 43.75             & \textbf{76.67}              \\ \midrule
\multicolumn{2}{c|}{Average}  & 44.38 & 52.26 & 59.24             & \textbf{75.70}      
     \\
\bottomrule

\end{tabular}}
\label{top25}
\end{table}

Common Weakness Enumeration (CWE)~\cite{CWE} is a list of vulnerability weakness types, which serves as a common language for describing and identifying vulnerabilities. The Top-25 most dangerous CWEs list officially publishes the most common and impactful software vulnerabilities over the previous two calendar years based on Common Vulnerabilities and Exposures (CVE) data~\cite{CWE-25}. Such weaknesses are dangerous compared to other vulnerabilities because they are more numerous and have a higher risk factor (CVSS).
In order to explore the effectiveness of \tool on the most common vulnerabilities,
we validate the effectiveness of \tool on the Top-25 most dangerous CWEs.


Specifically, we prepare the evaluation set by extracting the samples belonging to the Top-25 List from the Fan et al dataset. The evaluation set is named as \vd dataset for brevity in this paper.
The \vd dataset contains 8,989 code functions, and the specific type distribution is shown in Table 5. In terms of quantity, CWE-119 (the Bounds of a Memory Buffer)~\cite{CWE-119} and CWE-20 (Improper Input Validation)~\cite{CWE-20} have the largest proportion, with 28.61\% and 23.48\%, respectively.

We evaluate 12 types of vulnerabilities in the Fan et al.~\cite{fan} dataset. This is because only 12 of the 25 most threatening types of CWE appear more than 50 times. The remaining 13 data types appear too infrequently and are difficult to evaluate. We build a unified model for all types of vulnerabilities and use
a softmax function for testing based on the type of vulnerability. 
We compare with all the graph-based baselines that have been trained on the FFMPeg+Qemu model.
 
 As shown in Table~\ref{top25}, our \tool achieves more than 70\% accuracy on all vulnerabilities, with an average improvement of 
 27.78\%, compared to the previous best baselines. 
 It indicates that our method is able to discover different real-world vulnerabilities more accurately.
 Specifically, our method obtains the highest identification accuracy of 83.19\% on CWE-287 (Improper Authentication)~\cite{CWE-287} among all the types of vulnerabilities detected.
 In CWE-20 and CWE-119, which account for a large proportion, \tool has an accuracy of 75.32\% and 77.07\%, respectively, showing 30.13\% and 26.70\% improvement over
 previous methods, respectively.

 \begin{tcolorbox}
 \textbf{Answer to RQ3:} 
\tool achieves more than 70\% accuracy on real-world vulnerabilities, with an improvement of 27.78\% over the previous best-performing baseline.
 \end{tcolorbox}
 
 \subsection{RQ4: What is the influence of hyper-parameters on the performance of \tool?}
{In this section,} we explore the impact of {two} key hyper-parameters on the performance of \tool, including the number of layers of \gnn and the number of meta-path attention heads.
 
  \begin{figure}[t]
\centering
\subfloat[Layers]{\includegraphics[width=0.48\linewidth]{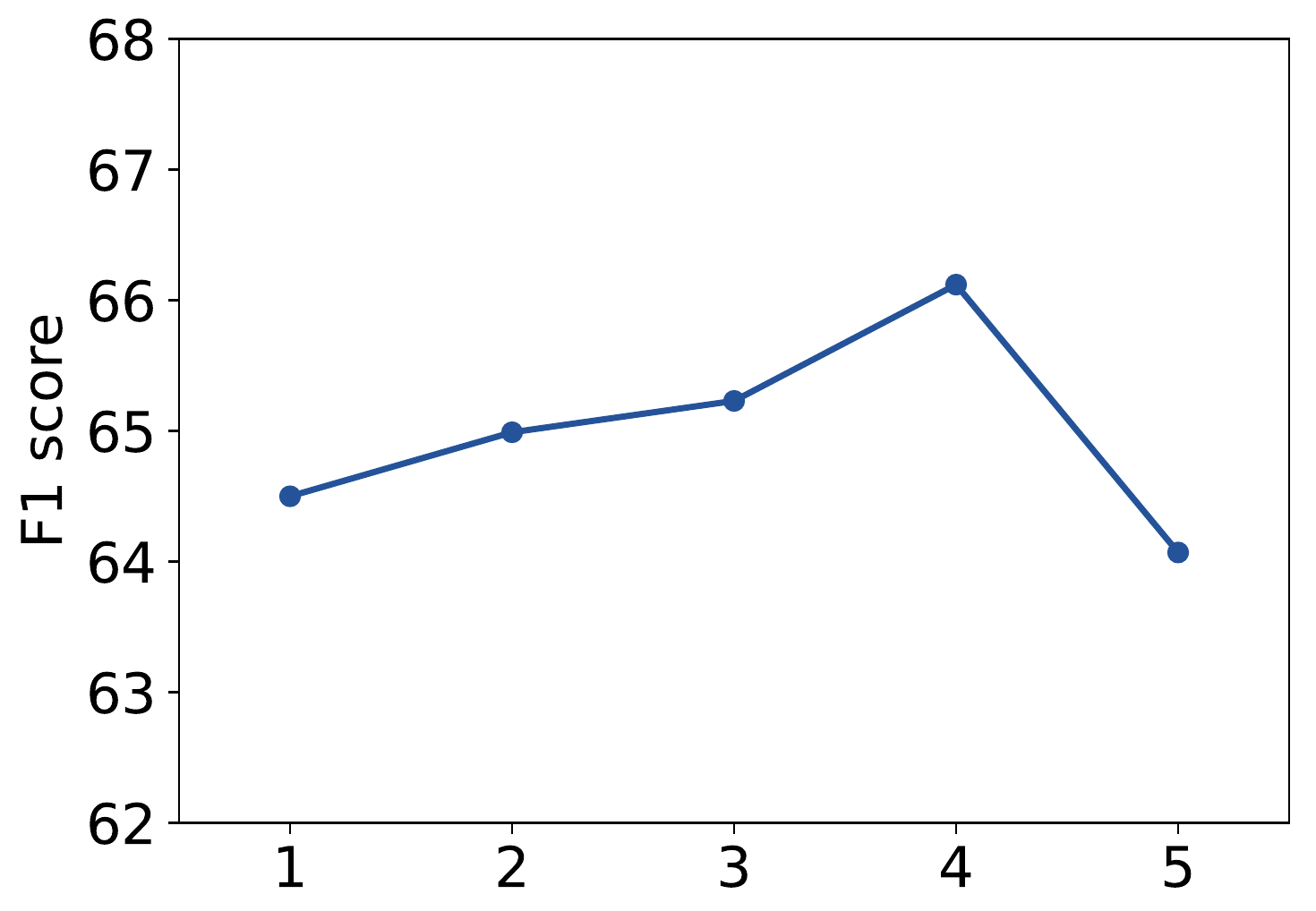}%
\label{layers}}
\hfil
\subfloat[Head numbers]{\includegraphics[width=0.48\linewidth]{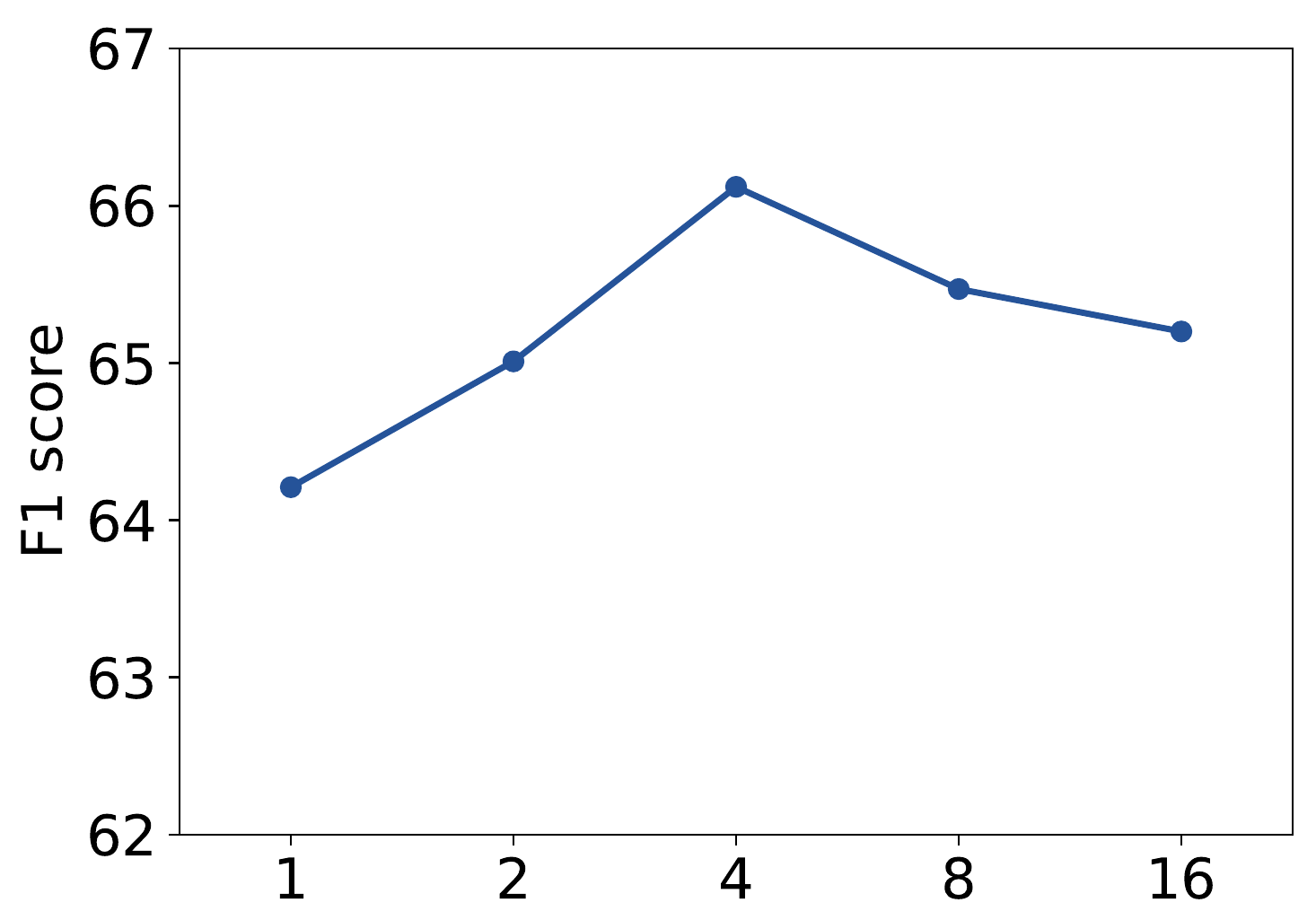}%
\label{headnumbers}}\\

\caption{Parameter analysis of \gnn's layers and head numbers of meta-path attention on the FFMPeg+Qemu dataset.}
\label{parameter}
\end{figure}

\begin{figure}[t]
\centering
\subfloat[Layers]{\includegraphics[width=0.47\linewidth]{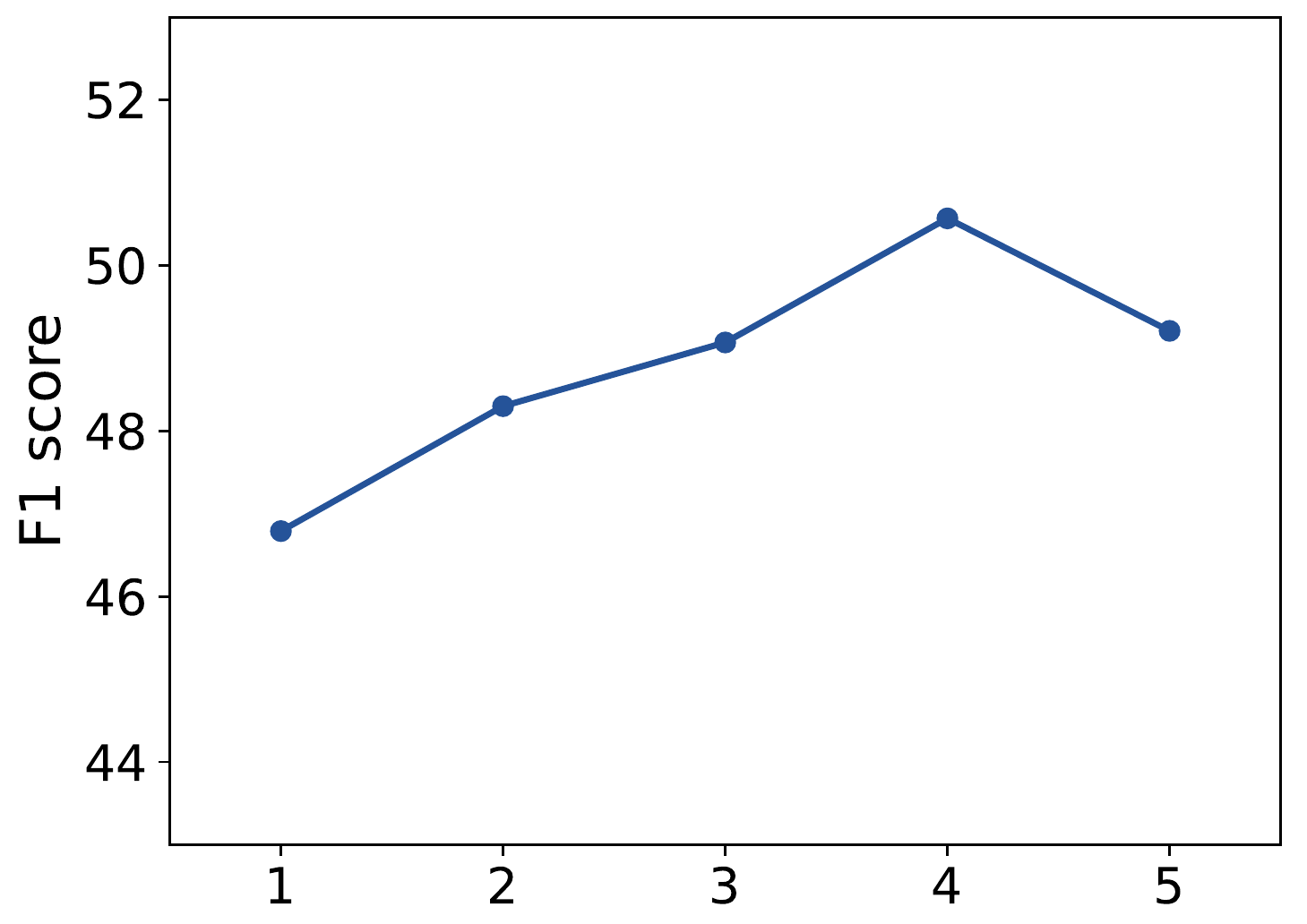}%
\label{layers_reveal}}
\hfil
\subfloat[Head numbers]{\includegraphics[width=0.47\linewidth]{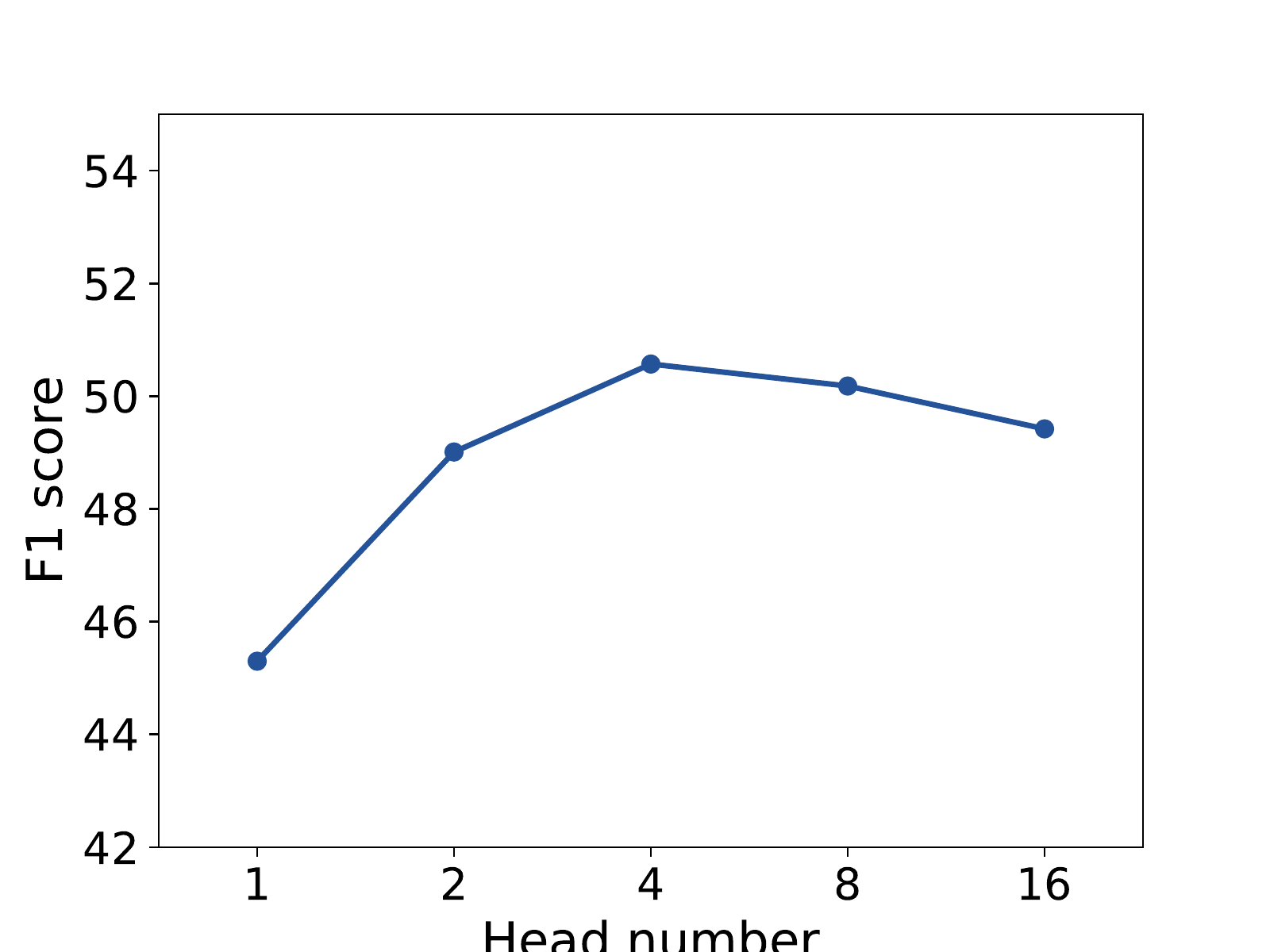}%
\label{headnumbers_reveal}}\\

\caption{Parameter analysis of \gnn's layers and head numbers of meta-path attention on the Reveal~\cite{reveal} dataset.}
\label{parameter_reveal}
\end{figure}

\begin{figure}[t]
\centering
\subfloat[Layers]{\includegraphics[width=0.47\linewidth]{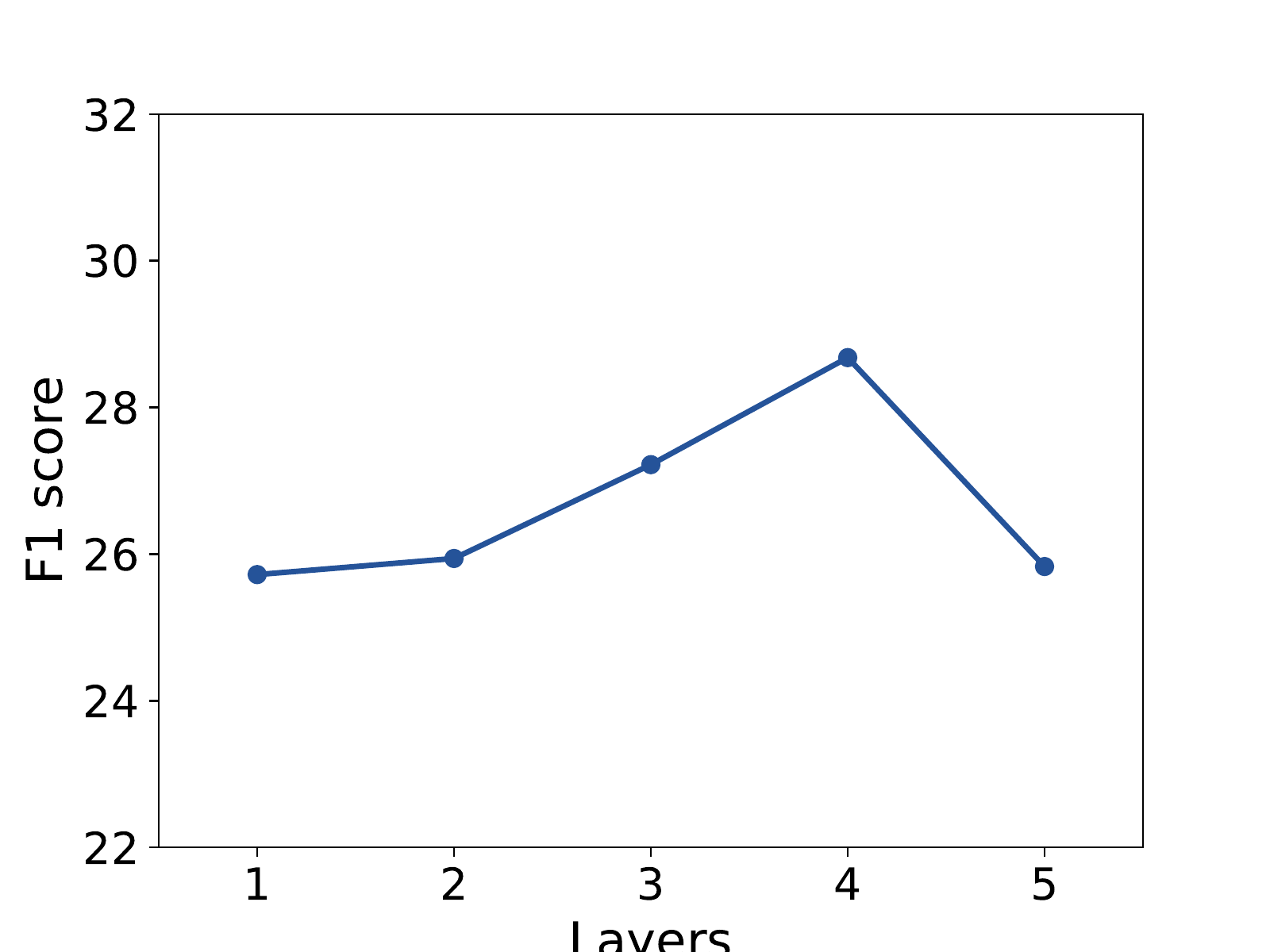}%
\label{layers_fan}}
\hfil
\subfloat[Head numbers]{\includegraphics[width=0.47\linewidth]{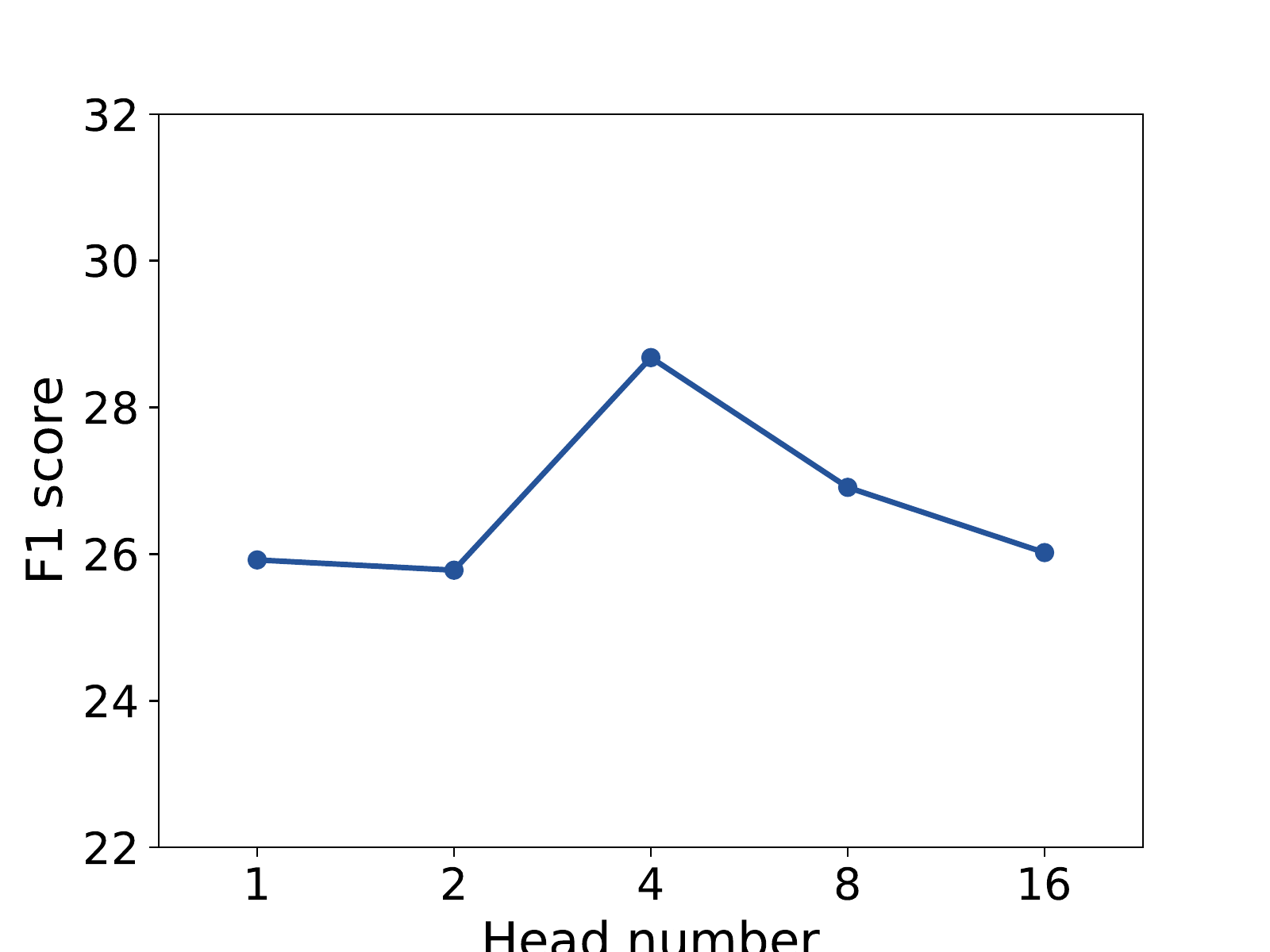}%
\label{headnumbers_fan}}\\

\caption{Parameter analysis of \gnn's layers and head numbers of meta-path attention on the Fan et al.~\cite{fan} dataset.}
\label{parameter_fan}
\end{figure}

\subsubsection{Layer Number of \gnn}

We explore the effect of different numbers of layers in \gnn on the performance of \tool on the FFMPeg+Qemu and Reveal datasets.
Fig.~\ref{layers} and Fig.~\ref{layers_reveal} show the F1 score of \tool with different numbers of \gnn layers. 
As can be seen, the F1 score of \tool first shows an increasing trend and then decreases as the number of layers increases on the FFMPeg+Qemu dataset, with a similar trend observed on the Reveal and Fan et al. datasets. The GNN layers are usually related to the distance between different nodes. The larger the distance between different nodes, the more message passing and GNN layers are required.


We find that \tool generally obtains the highest F1 scores when the number of layers is set as four, i.e., 66.12\%, 50.57\%, and 28.68\% on the FFMPeg+Qemu, Reveal and Fan et el. dataset, respectively.  We suppose that the \tool can better capture the information of the neighborhood as the layer number increases. However, as the layer number further increases, the over-smoothing issue would reduce the model performance.

\subsubsection{Attention Head Numbers}
We analyze the effect of different attention head numbers on the performance of \tool, with the results shown in Fig.~\ref{headnumbers}, Fig.~\ref{headnumbers_reveal}, and Fig.~\ref{headnumbers_fan}. As can be seen, \tool achieves the optimal F1 score when the number of attention heads is set as four. The trends of parameter setting in head numbers are roughly the same for all datasets.
Therefore, we empirically use four head numbers for all three datasets.

Overall, more heads show a significant improvement in F1 score compared to a smaller number of heads, which indicates that more heads are beneficial for capturing the code structure information in the meta-path graph. However, the performance starts to degrade after more than four heads.





 \begin{tcolorbox}
 \textbf{Answer to RQ4:} 

The hyper-parameter settings can impact the performance of \tool in the FFMPeg+Qemu, Reveal, and Fan et al. datasets. The experiment results show that both the head number and layer number are suggested to be set as four for achieving relatively better performance.
 \end{tcolorbox}

\section{Discussion}
\label{sec:discussion}

\subsection{Case Study}

\begin{figure}[t]
	\centering
    \includegraphics[width=0.49\textwidth]{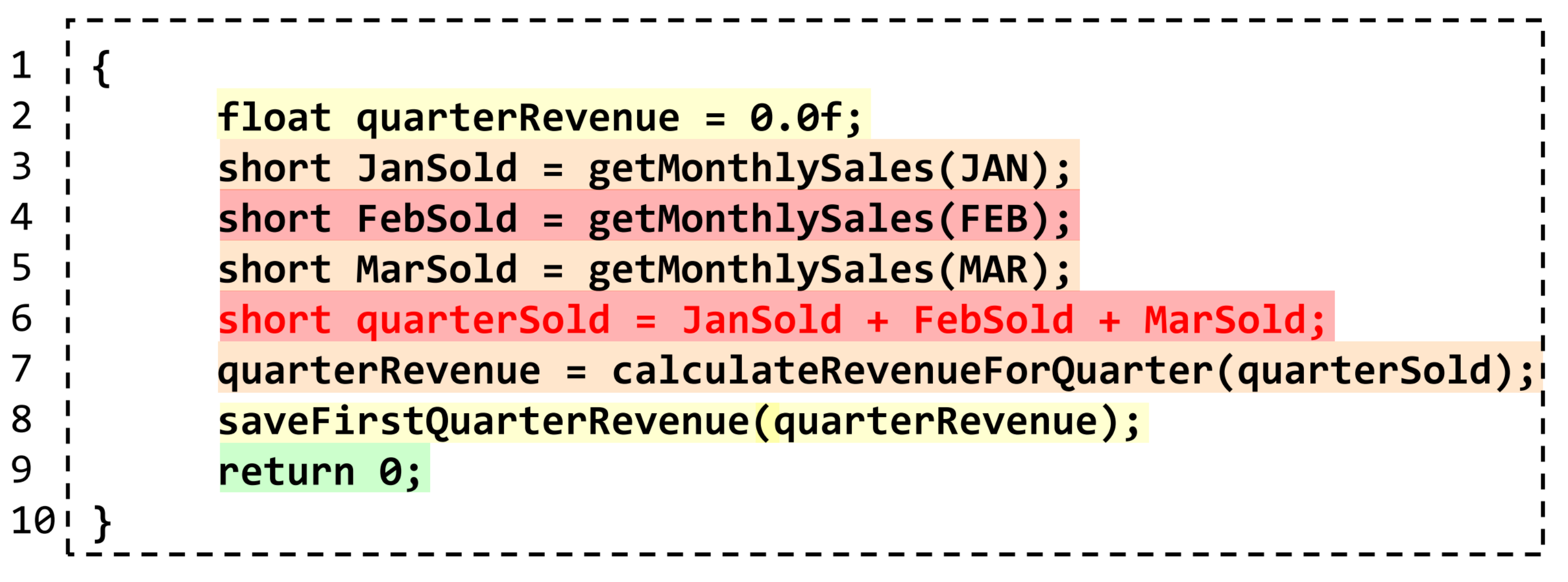}
	\caption{The heatmap of attention weights for a code example from CWE-190 (Integer Overflow or Wraparound). The code in red indicates a vulnerable statement. The red, orange, yellow and green-shaded statements indicate that the corresponding statements are associated with decreasing attention weights.
 }
	\label{caseexample}
\end{figure}

We conduct a case study to further verify the effectiveness of \tool in vulnerability detection. For analysis, we visualize the attention weight of each statement produced by \tool. Fig.~\ref{caseexample} visualizes the heatmap of attention weights for a vulnerable example from CWE-190.
In this example, the line 6 is the vulnerable statement, where the sum of the three variables may exceed the maximum value of the short int primitive type, producing a potential integer overflow. 
\tool notices the vulnerability 
and gives this statement the highest level of attention (red-shaded). The initialization statements (lines 3-5)
also present higher attention weights (orange-shaded and red-shaded). 
From the case, we guess that \tool is able to capture the code structural information and patterns of vulnerabilities, which is helpful for detecting code vulnerabilities effectively.

\subsection{Comparison with Pre-trained Models}

In recent years, there are some pre-trained approaches e.g., CodeBERT~\cite{DBLP:conf/emnlp/FengGTDFGS0LJZ20/codebert}, which aim at learning the general program representations. 
LineVul~\cite{fu2022linevul} leverages the CodeBERT model and selects vulnerability detection as the downstream task. And we compare the performance of \tool with LineVul. Specifically, we use the officially released code by LineVul and set default hyper-parameters of their paper.

We present a experimental results comparison between our tool (\tool) and LineVul, as summarized in Table~\ref{linevul}. 
The results indicate that \tool outperforms LineVul in terms of both Recall and F1 metrics on the FFMPeg+Qemu and Reveal datasets, with improvements of 53.92\% and 15.02\% respectively.  
This signifies that \tool is capable of detecting a greater number of vulnerabilities on these datasets compared to LineVul. On Fan et al. dataset, LineVul outperforms the \tool, which we believe may be due to the pre-trained model. However, LineVul uses the pre-trained model, which has additional data and contains more model parameters. Considering the scale of the data using (0.16M of \tool and 2.3M of LineVul) and the model parameters containing (0.65M of \tool and 125M of LineVul), we believe that \tool achieves comparable performance with LineVul.

\begin{table*}[h]
\centering

\setlength{\tabcolsep}{1.2mm}
\renewcommand{\arraystretch}{1.2}

\caption{Comparison results between \tool and the LineVul on the three datasets.}
\resizebox{.97\textwidth}{!}{
\begin{tabular}{l|cccc|cccc|cccc}
\toprule
{Dataset} & \multicolumn{4}{c|}{FFMPeg+Qemu \cite{devign}}        & \multicolumn{4}{c|}{Reveal \cite{reveal}}             & \multicolumn{4}{c}{Fan et al.~\cite{fan}}                \\
\midrule
Metrics                         & Accuracy & Precision & Recall & F1 score     & Accuracy & Precision & Recall & F1 score    & Accuracy & Precision & Recall & F1 score    \\
\midrule

LineVul                          & \textbf{64.75}& \textbf{63.98}& 50.51&56.45&
\textbf{92.43}& \textbf{48.86}&41.35&44.79&

\textbf{98.82}& \textbf{90.68}& \textbf{83.31}& \textbf{86.84} 

 \\

\tool
& 63.28   & 56.27    & \textbf{80.15} & \textbf{66.12} &91.60   &  42.86    &  \textbf{61.68} & \textbf{50.57} & 91.38   & 22.71  & 38.92 & 28.68\\
\bottomrule
\end{tabular}}

\label{linevul}
\end{table*}

\subsection{Performance on the Dataset Split by Time}

\begin{table}[h]
\centering

\setlength{\tabcolsep}{1.2mm}
\renewcommand{\arraystretch}{1.2}

\caption{Comparison results between \tool and the Reveal on the two settings of split in Fan et al.~\cite{fan}.}
\resizebox{.42\textwidth}{!}{
\begin{tabular}{l|cccc}

\toprule
Metrics     & Accuracy & Precision & Recall & F1 score  \\
\midrule

Reveal   & 77.39& 12.59& \textbf{25.73}&16.91 \\
\tool & \textbf{84.39}   & \textbf{17.01}    & {19.34} & \textbf{18.10} \\
\bottomrule
\end{tabular}}

\label{time_split}
\end{table}

Following the previous work~\cite{vuldeepecker, sysevr, devign, reveal, IVDETECT}, we have repeated 
three times of the experiment and taken the average to ensure the stability of the results. Furthermore, we employ a more rigorous data partition way to conduct experiments in Fan et al.~\cite{fan}, i.e., splitting data by time. Specifically, we use the ``update date" of the vulnerability data in Fan et al.'s dataset as the criterion for the data partition. Specifically,
we consider data updated before January 4th, 2018, as the training set, and the data updated
from January 4th, 2018, to January 15th, 2019, as the validation set. All samples beyond January 15th, 2019, were allocated to the test set.
We compared with the best-performing baseline Reveal to evaluate the influence of the time factor.
The detailed experimental results are shown in Table~\ref{time_split}.

It can be seen that based on the division of time, \tool gets three better performances out of a total of four cases. This illustrates that \tool is able to obtain better experimental results than the best-performing baseline when taking the time factor into consideration.  Specifically, \tool obtained 7\% and 4.42\% performance improvements in the Accuracy and Recall metrics, respectively. However, the dataset also suffers from data leakage. In the case of split by time, all four metrics degrade to different degrees. The degradation in the recall metric is the most severe for all methods, reaching an average of 12.96\%.

\subsection{Effectiveness of the Node Type Grouping Method in \tool}

\begin{table}[h]
\centering

\caption{ Performance comparison between four variants without node type abstraction in FFMPeg+Qemu~\cite{devign} datasets.}
\resizebox{.5\textwidth}{!}{
\begin{tabular}{l|c|cccc}
\toprule
\multicolumn{2}{c|}{Node type grouping}      & Accuracy                      & Precision                     & Recall                        & F1   
\\ \midrule
                                & Random    & 60.77                         & \textbf{57.73 }                        & 45.69                         & 51.01                         \\
\multirow{-2}{*}{5 node types}  & Frequency & 59.77                         & 54.80                         & 57.05                         & 55.90                         \\\midrule
                                & Random    & 60.88                         & 55.69                         & 61.05                         & 58.25                         \\ 
\multirow{-2}{*}{10 node types} & Frequency & 61.89                         & 55.66                         & 72.41                         & 62.94                         \\
\midrule
\multicolumn{2}{c|}{\tool}    & \textbf{63.28} & 56.27 & \textbf{80.15} & \textbf{66.12} \\
\bottomrule
\end{tabular}}
\label{withoutabs}
\end{table}

Using the code structure graph without abstraction in \tool is cost intensive. For example, the data preprocessing steps of the FFMPeg+Qemu~\cite{devign}, Reveal~\cite{reveal} and Fan et al.~\cite{fan} required approximately 17, 9, and 170 hours on NVIDIA GeForce RTX 3090 GPU, respectively.
The expenditure of resources and time is a consequence of the excessive number of meta-paths. The code structure graph consists of 69 node types and 4 edge types, resulting in a total of $69\times 4\times 69 = 19044$ Node-Edge-Node meta-paths. This profusion of meta-paths can lead to
the out of CUDA memory issue for model training. To validate the effectiveness of our proposed node type grouping, we generate the code structure graph of node types with other abstraction methods. 

Specifically, we use two methods to keep $q$ node types: based on the number of nodes and random selection. We use $q$ for 5 and 10 different node types, resulting in four node type abstraction methods.
The maximum number of node types is set as 10, since it results in 400 meta-paths, which is close to
maximum CUDA memory capacity of the GPU. 
The experiments are performed the FFMPeg+Qemu dataset for evaluation. The preprocessing step takes approximately three hours, three times longer than that
of our \tool. 
The results also validate the effectiveness of the node type grouping strategy in \tool.
As shown in Table~\ref{withoutabs}, \tool outperforms all the four variant approaches, achieving
average improvements of 2.45\%, 0.3\%, 21.1\%, and 9.10\% for the accuracy, precision, recall, and F1 score, respectively. 
Furthermore, the selection
strategy based on node frequency showcases notable performance improvements compared with the random selection strategy.
\subsection{Effectiveness of Meta-path}

In vulnerability detection, the
code structure graph is also a heterogeneous graph. Constructing the meta-path in the heterogeneous network can capture the vulnerability-related information among nodes in the code structure graph. 
\tool can adjust the different weights of the meta-path, causing the model to focus its attention on specific paths.

Specifically, in Fig.~\ref{longexample}, we denote
three meta-paths associated with a statement and its sub-nodes, represented by meta-paths A, B, and C, respectively. Traditional vulnerability detection methods, such as Devign and Reveal, assign equal weights to all the paths, requiring the model to process a large amount of information for learning. 
This tends to result in inaccurate vulnerability detection.
In this case, meta-path C is the most critical in terms of the vulnerability trigger path, as the assignment relations between nodes referred to by NCS determine the vulnerability pattern. After model training, the weight assigned to meta-path C is 0.44, while the weights for the other paths are ignorable
($-9.09E^{-31}$ for the
meta-path A and $-1.99E^{-30}$ for the
meta-path B). It indicates that the model effectively disregards the less relevant paths and focuses on the more critical part of the vulnerability.

\subsection{Threat to Validity}
\textit{Dataset Partition.}
None of the existing baseline methods publish their divisions of datasets, so we can not completely reproduce the previous results. Following the data division of previous work~\cite{devign, reveal}, we perform a dataset division and reproduce the experimental results based on their articles as much as possible.
Reveal and IVDetect use different preprocessing methods, which may lead to inconsistencies in the dataset under the same division. We relied on their source code for the preprocessing work as much as possible. Devign~\cite{reveal} did not publish the source code of their implementation, we reproduce Devign based on Reveal's~\cite{reveal} implementation version and try to be consistent with the original description.



\textit{Generalizability of Other Programming Languages.} Our node classification is based on the AST node type in C/C++. Therefore, we only conduct experiments on the C/C++ dataset and do not choose other programming languages such as Java and python. However, the main idea of \tool can be generalized to other programming languages because the approach does not rely
on language-specific features. We will
evaluate \tool on more programming languages in our future work.


\section{Related Work}
\label{sec:related}

Recently, learning-based vulnerability detection has been a significant research problem in software engineering. 
Depending on how the source code is represented and which type of learning model is utilised, existing technologies can be generally divided into two different types: token-based and graph-based methods. 

The token-based methods~\cite{vuldeepecker, russell, sysevr, DBLP:conf/iclr/AlonBLY19/code2seq, DBLP:conf/ijcnn/NguyenLVMGP21/ICVH} treat the code as a sequence of tokens, which contain two phases: feature extraction and training. In the feature extraction phase, the token-based methods usually  extract token-based features as the model input, which include identifiers, keywords, separators, and operators. These features are usually specified and written by developers and can represent the different line structure~\cite{DBLP:journals/tse/KamiyaKI02} of the code. For example, Russell \et~\cite{russell} divides each code fragment at the function level and treats them as an individual sample. It generates a lexical token sequence for each function to represent the whole code sample feature set. Code2Seq~\cite{DBLP:conf/iclr/AlonBLY19/code2seq} uses path-context extracted features for each program method and splits code tokens into subtokens. In the training phase, these methods treat source code as sequences via utilizing various deep neural networks. Russell \et use CNN~\cite{DBLP:conf/emnlp/Kim14} for code vulnerability detection, concatenating these token features through convolution filters. SyseVR~\cite{sysevr} uses GRU~\cite{DBLP:conf/emnlp/TangQL15/gru2} to capture the sequence information of the code. Code2Seq uses BiLSTM~\cite{DBLP:journals/nn/GravesS05/lstm} to encode the necessary information in a sequence of tokens. 


In recent years, the graph-based methods~\cite{devign,reveal,IVDETECT,DBLP:conf/iceccs/ChengWHZXYS19/VGDETECTOR,DBLP:conf/wcre/DingSZLMKR22/VELVET} have achieved state-of-the-art performance on vulnerability detection. They capture more structural information in the source code than token-based methods. They generally represent source code snippets as graphs generated from static analysis.
Depending on different code representations, they design various GNN models for detecting code vulnerabilities.
For example, VGDETECTOR~\cite{DBLP:conf/iceccs/ChengWHZXYS19/VGDETECTOR} uses CFGs to embed the execution order of a code sample and then uses the GCN~\cite{kipf2016gcn} to capture neighborhood information in the graph structure. Reveal~\cite{reveal} first generate CPG~\cite{cpg} and use Word2Vec~\cite{DBLP:journals/nle/Church17/word2vec} to initial the node vector representations of the code tokens. Then they use GGNN~\cite{ggnn} to detect code vulnerabilities. 


Li et al. \cite{DBLP:journals/pacmpl/LiWNN19} propose a bug detection approach that utilizes Program Dependency Graph (PDG) and Data Flow Graph (DFG) as the global context, while leveraging extracted information from the Abstract Syntax Tree (AST) as the local context. They treat the AST as a homogeneous graph, focusing solely on the relationships between different node vectors along paths. The difference between Li et al.'s work and \tool is the targeted programming languages. The parsing process in Li et al.'s approach is specifically designed for the Java programming language, which is hard to be applied to C/C++ language.

However, all these methods focus on learning the local features of nodes and fail to capture heterogeneous relations and long-range dependencies among different types of nodes and edges. For example, 
Devign~\cite{devign} treats all nodes as the same node type and only uses the node value in the code structure graph, which leads to missing information on the node type and ignoring heterogeneous relations in the code structure graph. Cao et al.~\cite{cao2022mvd} construct the Program Dependence Graph (PDG) instead of a code structure graph to capture the interprocedural flow information including data flow and control flow, which only focus on the statement nodes and do not consider the other nodes. 
In this paper, we propose a meta-path based attentional graph learning model to learn the heterogeneous relations and long-range dependencies in the code structure graph.

\section{Conclusion}
\label{sec:conclusion}
In this paper, we propose \tool, a meta-path based attentional graph learning
model for vulnerability detection. \tool consists of a multi-granularity meta-path construction, which consider the heterogeneous relations between the different node and edge type. We also propose a multi-level attentional graph neural network \gnn to comprehensively capture long-range dependencies and structural information in the meta-path graph. 
Our experimental results on three popular datasets validate the effectiveness of \tool, and the ablation studies and visualizations further confirm the advantages of \tool.
Compared with state-of-the-art deep learning-based methods, \tool gains better performance and detects more vulnerabilities in the real world.
The implementation of \tool and the experimental data are publicly available at: \textit{{\http}}.

\bibliographystyle{IEEEtran}
\end{document}